\documentstyle[aps,psfig]{revtex}
\begin{document}
\def\ybco{YBa$_2$Cu$_3$O$_{7-\delta}\, $}
\def\bscco{Bi$_2$SrCa$_2$Cu$_2$O$_8\, $}
\title{Nature of the Low Field Transition in the Mixed State of
High Temperature Superconductors}
\author{Seungoh Ryu\cite{presentaddress} and David Stroud}
\address{Department of Physics, Ohio State University, Columbus, OH 43210}
\date {\today}
\maketitle

\begin{abstract}

We have numerically studied the statics and dynamics of a model
three-dimensional vortex lattice at low magnetic fields.
For the statics we use
a frustrated $3D$ $XY$ model on a stacked triangular lattice.
We model the dynamics as a coupled network
of overdamped resistively-shunted Josephson junctions with Langevin noise.
At low fields, there is a weakly first-order
phase transition, at which the vortex lattice melts into a line liquid.
Phase coherence parallel
to the field persists until a sharp crossover, conceivably a phase transition,
near $T_{\ell} > T_m$ which develops at the same temperature as an 
{\em infinite} vortex tangle.
The calculated flux flow resistivity in various geometries near $T=T_{\ell}$
closely resembles experiment.
The local density of field induced vortices increases sharply
near $T_\ell$, corresponding to the experimentally observed magnetization
jump.  We discuss the nature of a possible transition or crossover at $T_\ell$(B) 
which is distinct from flux lattice melting.
\end{abstract}

\section{Introduction}
Ever since their discovery, the behavior of high-T$_c$ materials in a magnetic
field has seemed mysterious\cite{malozemoff89}.
Unlike the conventional low-T$_c$ type-II materials,
high-T$_c$ superconductors (HTSC's) show a broad region in the
magnetic-field/temperature (H-T) plane where the Abrikosov lattice has
apparently melted into a {\em liquid} state\cite{nelson88}.

Considerable recent evidence
now suggests that flux lattice (FL) melting is
a {\em first-order} phase transition.
On the experimental side, a local magnetization jump
has been measured by network
of Hall microprobes\cite{zeldov95} on \bscco and has been
associated with the melting transition.
The transition thus observed seems to lie quite near the
melting curve as determined from low angle neutron
diffraction\cite{cubitt93} and $\mu$SR experiments\cite{lee93}.
More recently, Schilling {\it et al}\cite{schilling96,schilling97} have
directly observed the latent heat of the transition
in \ybco along a line $T_m(H)$ in the $H-T$ phase diagram which 
agrees well with mechanical
and transport measurements\cite{safar92,farrell91,farrell95}.
Numerical evidence for a first order melting
transition has been obtained from simulations based on a frustrated $XY$
model\cite{hetzel92,dominguez95},
and from a lowest Landau level model which is expected to be most accurate
at high magnetic field\cite{sasik95}.  First order melting has also been
observed numerically
in a system of unbreakable flux lines described by a
Lawrence-Doniach model\cite{ryu96}. All these simulations are based on
a large density of flux lines ($\sim {\cal O}(1-10) {\rm tesla}$).

An anomalous feature of the local Hall probe measurements is that the
apparent first order transition line seems to terminate at a critical point
above which the latent heat
vanishes\cite{zeldov95}.
Since on symmetry grounds
a first order ``melting'' line cannot terminate in a critical
point\cite{landau}, this critical point may suggest
that the first order melting line is instead intersecting another
phase transition line related to the disorder.
A related issue is the entropy
released per vortex per layer across the transition line.  This entropy
increases very rapidly as the field decreases.  Such behavior is difficult to
accounted for within a model based only on the field induced vortices.

In the presence of disorder, the lattice becomes unstable at high fields
against proliferation of quenched-in topological defects\cite{ryu96_2,gingras96},
possibly through a first order phase transition across a horizontal (constant
H) line in the H-T plane.   This line then may meet the
temperature-driven melting line, causing it to terminate.
Somewhere along this line,
the melting transition may be converted
into the universality class of the continuous vortex
glass transition\cite{fisher89}, characterized by divergent correlation
lengths and times.

Another unresolved issue regarding the phase diagram is the
possibility of reentrant melting at low fields.   Reentrant flux lattice
melting is expected because of screening of the widely-separated vortex
lines at low fields\cite{nelson88,ryu92}.  It
has been recently reported in single-crystal
${\rm NbSe_2}$ sample\cite{sabu}, based on tracking of the 
so-called ``peak effect''\cite{kwok94,ling95}.  
Such reentrance behavior has been observed only at the limited field range by Ling {\em et
al}\cite{ling97}.
On the other hand,  the melting curve tracked by the micro-Hall
probe\cite{zeldov95}  seems to monotonically approach 
the zero-field superconducting transition at $T_c(H=0)$ even for fields
as low as a few Gauss.

FL melting can also be probed by transport measurement.  But since such
measurements are non-equilibrium, they offer only an indirect means of
studying {\em equilibrium} FL melting.  In real materials with disorder, the
interpretation of transport measurement is further complicated by the
many competing energy scales.  In single-crystal \ybco, 
the in-plane resistivity exhibits a
discontinuous jump and hysteresis which have been identified with a
first-order melting transition\cite{safar92}.  Nonetheless, the
peak effect in the critical current
occurs at slightly lower temperatures than the
resistivity jump, leading some workers to postulate that there
is a ``premelting'' phenomenon\cite{kwok94} in this material,
in addition to melting.   In \bscco,
simultaneous transport and local magnetization measurements\cite{fuchs96}
show that the jump in local magnetization $M$ coincides
(at high fields) with a jump in the resistivity $\rho_{ab}$
from zero to a finite value, or (in low fields) the continuous development
of a finite $\rho_{ab}$.   In addition, at high fields, 
the jumps in $\rho$ and $M$ are
accompanied by hysteresis.  Together, these phenomena strongly suggest 
first order flux lattice melting at high fields.   At low fields, the 
experiments are more ambiguous.

FL melting has been widely studied numerically.  
The possibility of {\em two stage} melting
was first suggested by Li and  Teitel\cite{li93_1,li93_2}
for a model with infinite penetration depth $\lambda$, and later
for a system with finite $\lambda$\cite{chen95,chen97}.  The
calculations of Li and Teitel are based
on the so-called frustrated XY model with fairly low flux per plaquette
of $f = 1/25$ (in units of the flux quantum $\Phi_0=hc/2e$) on a simple cubic 
lattice. 
They find that the three-dimensional flux line
lattice (3D FLL) melts first into a ``line liquid''
characterized by disentangled flux lines,
which become entangled at a second, higher-$T$ phase transition.
Current-voltage (IV) measurements in the so-called ``flux transformer''
geometry\cite{safar94,lopez94,keener96,keener97} provide some support for
this picture.  Specifically, they suggest that FL melting is signaled by
the onset of finite in-plane resistance, while in an applied current,
phase coherence is lost in the $c$ direction only at a distinctly higher
temperature.
On the other hand, simulations of dense (f=1/6)
flux lines on a stacked triangular
grid favor a single transition\cite{hetzel92,dominguez95}.  
Dynamical calculations\cite{jagla96} on a triangular lattice at
$f = 1/6$ suggest that if there are two separate transitions, they arise from pins,
either intrinsic to the discrete cubic grid, or put in by
hand. Yet more recent studies based on
a London vortex loop model on a simple cubic lattice
show that superconducting order disappears apparently in two steps, 
the sequence of which depends on the
lattice anisotropy\cite{nguyen96}, although it is argued to be a 
finite size effect by the same authors\cite{nguyen97}.

In this paper, we attempt to resolve some of these issues by considering
the frustrated $XY$ model over a {\em wide range of flux densities},
using both static and dynamic simulations but with no quenched disorder.
By examining this model on a stacked triangular lattice,
we minimize the unphysical periodic pinning due to the lattice.  By
working at relatively low densities, we focus on the regime, now being
probed experimentally, where the XY phase fluctuations (vortex loops) are as
important as those of {\em field induced} vortex lines\cite{tesanovic95}.
Our main conclusion is that there are, in fact
signatures of two separate transitions at low fields,
which are not artifacts of pinning by the discrete grid.
The transition at lower temperature is unambiguously associated 
with vortex lattice melting.
The second transition may be a sharp crossover rather than a true phase
transition.  Nevertheless, it is responsible for several
experimental features (such as sharp increases in local magnetization and in 
resistance) which are often identified as evidence for a first order
melting transition.

The remainder of this paper is organized as follows.
In Section \ref{modelsec}, we
describe our model and its numerical solution.
The following sections
present our numerical results,
which are followed by a discussion and then summarized in a concluding
section.

\section{Model}
\label{modelsec}

\subsection{Hamiltonian and Thermodynamics}

We study the standard frustrated $XY$ model described by the Hamiltonian
\begin{equation}
\label{modeleq}
{\cal H} = -J\sum_{\langle ij \rangle}\cos(\theta_i-\theta_j -
A_{ij}),
\end{equation}
where
$A_{ij} = \frac{2\pi}{\Phi_0}\int_{i}^j{\bf A}\cdot{\bf dl}$,
${\bf A}$ is the vector potential
associated with a uniform magnetic field ${\bf B} = B\hat{z}$ applied parallel
to $\hat{z}$, $\Phi_0 = hc/2e$ is the flux quantum,
$\theta_i$ is the phase of the order parameter on site
$i$, and the sum runs over nearest neighbor pairs.
We use a stacked triangular grid with ${\bf B}\| z$, the direction
perpendicular to the triangular network, with periodic boundary conditions
(PBC) in all directions except where stated otherwise.

To allow a wider range of frustrations compatible with
the boundary conditions, we use a variant of the Landau gauge\cite{haldane}.
Note that there are four bonds per grain: three in the xy-plane and one along
$\hat{z}$.
We label these by their unit vectors ${\hat{\alpha} = \hat{x},
\hat{y_1},
\hat{y_2},
\hat{z}}$ where
$\hat{y_1} =(1/2)\hat{x} + (\sqrt{3}/2) \hat{y}$ and
$\hat{y_2} = -(\sqrt{3}/2)\hat{x} + (1/2) \hat{y}$.
The phase factors $A_{ij}$ connecting a grain located at (x,y,z) to
its four nearest neighbors are given by
$0$ along $\hat{x}$ or $\hat{z}$,
$2 \pi f  (2 x + 1/2)$ along $\hat{y_1}$, and $2\pi f (2 x - 1/2)$
along $\hat{y_2}$.  There are exceptions to this form for grains
lying on the boundaries:  All grains lying on the $x = {\rm L}_x$ boundary plane
have
$A_{ij} = - 2\pi \cdot 2 {\rm L}_x y$ for bonds in the $x$ direction.
Bonds at $x={\rm L}_x$ boundary such that ${\rm mod}(j,2) = 1$ have
$A_{ij} = 2 \pi f [2 x + 1/2  - 2 {\rm L}_x ( y +1 ) ] $ in the $y_1$ direction.
For bonds on the $x=0$ boundary with ${\rm mod}(j,2) = 0$,
$A_{ij} = 2 \pi f [ 2 x - 1/2 + 2 {\rm L}_x (y+1) ] $ for
bonds in the $y_2$ direction.   In contrast to the usual Landau gauge (which
is compatible with frustrations only in integer multiples of $1/(2N_x)$), this
generalized gauge is compatible with any $f$ which is an integer multiple
of $1 / (2N_xN_y) $ under periodic boundary conditions.

We have considered networks of sizes ${\cal N} = N_x \times N_y \times N_z$.
For $f=1/24$, we have studied
$N_z = 12, 24, 48$ and $N_x = N_y = 24$, and for other values of
$f$ (1/2592, 1/1648, 1/81, and 1/6) we have considered $N_x=N_y=N_z=18$.
In two dimensions, the vortices lie on the vertices of a honeycomb grid of
unit length of $(1/\sqrt{3})a_B$ which is dual to the
triangular grid of unit length $a_B$. Assuming that vortices form perfect
triangular lattice on this grid, and equating the area per vortex
to $(\sqrt{3}/4) a_B^2 /f,$ we obtain the following necessary
condition for a triangular
vortex lattice to form without {\em geometric frustration} of the FL:$
2/f = (n_1^2 / 3 + n_2^2 / 4)$ with integers $n_1, n_2$.
The values of $1/f$ satisfying this condition are then
$2,6,8,14,18,24,32,38,42 \ldots, 648,
\ldots$. For $f=1/162,$ two distinct pairs [$(n_1,n_2)$ = (0,36), and (27,18)]
satisfy the condition.
$f=1/648$ has the lowest possible nominal vortex density of one per $18 \times
18$ system compatible with our chosen gauge, and allows either of the
pairs $(n_1,n_2)$ = (0,72) and (54,36).
$f=1/2592 ( = 1/4 \times {1 \over 2\times18\times18}),$ represents less
than a single vortex line, and the system is {\em gauge-frustrated}.

In practice, for $f \leq 1/81$, there are too few
field-induced flux lines to study FL melting.
Nonetheless, the dilute regime is still of interest, since in these cases,
the flux lines behave independently and the
thermodynamics is
dominated by the zero-field phase degrees of freedom\cite{tesanovic95}.
Except for f = 1/2592, we study only gauge-unfrustrated values allowing 
only an integer number  of vortices in the simulation box.

We calculate the thermodynamics using
a standard Monte Carlo (MC) algorithm, with up to $10^6$ MC steps
at each temperature $T$. To ensure equilibration in the ground state
for all values of $f$, we performed simulated annealing
runs for the two-dimensional (2D) version at each $f$ with the same lateral
dimensions.  We then form the
ground state of the 3D system by stacking the 2D ground states thus found
uniformly along the $z-$direction.
This enables us to find the ground state configuration of a perfect triangular
lattice for low values of $1/24 \le f \le 1/18$  within a reasonable
time.  Starting from these 3D ground states, we warm up the system in
steps of
$\triangle T / J = 0.05$ or $0.1$,
allowing at least $4-5 \times 10^4$ Monte Carlo sweeps for each $T$.
The final configuration for each $T$ is then saved to be used as
a starting configuration in some of the longer calculations as well in the
dynamic simulations.

From these calculations, we extract a range of thermodynamic quantities.
One of these is the specific heat
$C_V = (\langle H^2 \rangle - \langle H\rangle^2)/(k_BT)$ at temperature $T$.
We also calculate the local vorticity vector field $n_\alpha(p)$ defined for
each Cartesian direction $\alpha$ and each point $p$ of the stacked honeycomb
dual lattice.  At each instant during the simulation,
$n_\alpha(p)$ is determined from
\begin{equation}
	\sum^\alpha_p \rm{mod} [ \phi_i - \phi_j - A_{ij}, 2\pi ] =
2 \pi [ n_\alpha(p)  - f_p ].
	\label{vortdefeq}
\end{equation}
Here the summation runs along the bonds belonging to the plaquette labeled
$p, \alpha$ (a triangle in the $xy$ plane, a square in planes parallel to the
$z$ axis); and $f_p
\equiv \sum^\alpha_p A_{ij}/(2\pi).$
From $n_{\alpha}$,
one can also compute the {\em structure factor},
$S_{\alpha\beta}({\bf k}) = \langle n_{\alpha}({\bf k})n_{\beta}(-{\bf
k})\rangle$, where $n_{\alpha}({\bf k})$ is the Fourier transform of the
local vorticity vector n$_{\alpha}({\bf r})$.
We also calculate the principal components $\gamma_{xx}$ and $\gamma_{zz}$ of
the helicity modulus tensor\cite{fisher73}, in the directions perpendicular
and parallel to the applied field.
To within a constant factor, ${\bf \gamma}$ represents the
phase rigidity or the superfluid density tensor of the system; its
derivation in
terms of equilibrium thermodynamic averages has been given
elsewhere\cite{shih84}.

\subsection{Dynamics}

To treat the dynamics, we model each link between grains as
an overdamped resistively shunted Josephson junction (RSJ)
with critical current $I_c = 2eJ/\hbar$, shunt
resistance $R$, and Langevin white noise to simulate temperature effects.
The effective $IV$ characteristics are then obtained by numerically integrating
the coupled $RSJ$ equations, as described elsewhere\cite{chung89}, using a time
constant typically of $0.1 t_0$ and obtaining voltages by averaging over an
interval of $\sim 600 t_0 - 2000 t_0$.
Since direct solution of these equations would involve inverting
and storing an ${\cal N}\times {\cal N}$ matrix, where ${\cal N}$ is the
total number of grains$ \sim  {\cal O}(5000),$ we instead solve them
iteratively\cite{recipe}, incurring a speed penalty of a factor of
$\ln {\cal N}.$  We verified that our solutions converge by comparing
them with those from direct inversion for time steps of
$0.01, 0.04$ and $0.1$ on an $8 \times 8 \times 8$ system.

The most obvious approach to the dynamics of this model would be to
use free boundary conditions, injecting current into one face of the
lattice and extracting it from the opposite, with periodic transverse
boundary conditions\cite{lee93a}.  But this has the following
disadvantage.  Once the flux lattice is depinned from its underlying periodic
pinning potential, it will drift along {\em as a whole} under the influence of
the Lorentz force provided by the driving current.  Since this occurs
equally in the solid and the liquid state, such a geometry may not distinguish
clearly between flux lattice and flux liquid (in the absence of spatially
inhomogeneous pinning centers).  This problem may be even more
conspicuous in our stacked triangular geometry, since the critical current
$I_{dp}$ for depinning a single vortex pancake from
the underlying triangular grid at zero temperature (T = 0) is smaller
($I_{dp} \approx 0.037I_c$) than in a square grid
($I_{dp} \approx 0.1I_c$)\cite{lobb}.

\begin{figure}[th]
\begin{center}\leavevmode\psfig{figure=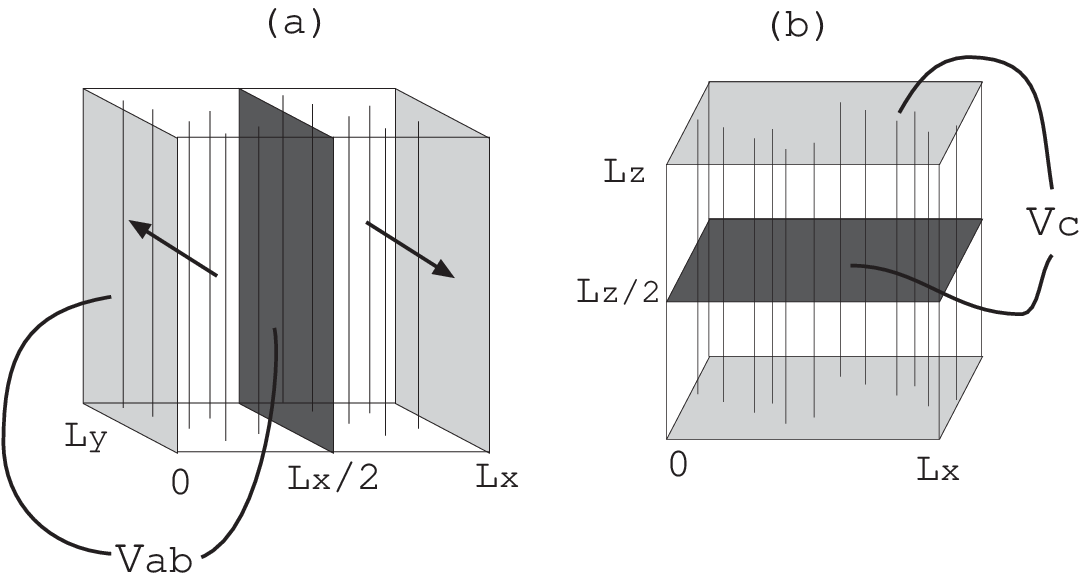,width=4in}\end{center}
\caption [lala] {Geometry for the dynamic calculations described in the text.
(a) To probe the shear modulus, current is injected uniformly into plane
$\Sigma_{in}$
and extracted uniformly from $\Sigma_{out}.$
The arrows indicate directions of the Lorentz forces acting on
the lines in the two half volumes (in opposite directions, because of the
periodic boundary conditions).
(b) To probe the c-axis resistance, the currents are injected and extracted
uniformly from the two planes
indicated; the voltage drop between the planes is measured
as in (a).
}
\label{geomfig}
\end{figure}

We therefore adopt a different geometry for injecting and extracting current,
as shown in Fig.\ \ref{geomfig}.  Fig.\ \ref{geomfig} (a) corresponds to
injecting current $I/I_c$ into each grain in the $yz$ plane at x = 0, and
extracting it from each grain at $x = N_x/2$ (with periodic boundary
conditions in all three directions).  In this geometry, the Lorentz forces
acting on the vortices in the two halves of the volume are oppositely directed,
as indicated by the arrows.  Thus, in this geometry, we are effectively probing
the {\em shear modulus} $\mu$ of the vortex lattice, on a length scale
${\rm L}_x/2$.  Similar geometries have been previously discussed in the context
of possible experiments\cite{pastoriza95,nelson91}.  In Fig.\ \ref{geomfig}
(b), we show a geometry which is designed to probe
the {\em c-axis resistivity}, $\rho_c$.  In this case, we inject a current
$I$ into each grain on the $xy$ plane at $z = 0$ and extract it from each
grain at $z = N_z/2$.  There is on average no Lorentz force on
the vortex lines.

\section{Zero-Field XY Model: f = 0}

\subsection{Thermodynamics}

The 3D unfrustrated XY model on a {\em cubic grid} has been
extensively studied\cite{gottlob93}.  Near the phase transition,
the specific heat $C_V \sim |T - T_{XY}|^{-\alpha}$
with $\alpha \sim 0$ and the correlation length
$\xi \sim |T - T_{XY}|^{-\nu}$ with
$\nu \sim 0.66-0.67$.  For a cubic lattice,
$T_{XY} \sim 2.203 J$.
In a stacked triangular grid, where each grain has more nearest neighbors,
the transition  is shifted to a higher temperature.
Numerically, we find that $T_{XY} \sim 3.04 J$.

The $XY$ phase transition is best characterized by the
{\em helicity modulus tensor} $\gamma_{\alpha\beta}$, which measures
the phase rigidity of the system\cite{fisher73}.
In stacked triangular lattice, this tensor is diagonal with elements
\begin{eqnarray}
	\gamma_{\alpha\alpha} &=& {1\over V} {\delta^2 {\cal F} \over
\delta A_{\alpha}^*\delta A_{\alpha}^* }
	\nonumber \\ &=& {J \over V} \Big< \sum_{ij} \cos(\Theta_{ij})
\hat{n}_{ij}\cdot
	\hat{n}_\alpha \Big> - {J \over k_B T} \times \nonumber \\
	& & \Big( \Big< [ \sum_{ij}\sin(\Theta_{ij})\hat{n}_{ij}\cdot
	\hat{n}_\alpha ]^2 \Big> - \Big<
\sum_{ij}\sin(\Theta_{ij})\hat{n}_{ij}\cdot
	\hat{n}_\alpha  \Big>^2 \Big).
	\label{heldefeq}
\end{eqnarray}
Here $A^*_{\alpha}$ is a fictitious uniform vector potential in the $\alpha$
direction, $V$ is the volume, $\Theta_{ij}=\theta_i- \theta_j - A_{ij}$
is the gauge-invariant phase difference, $\hat{n}_{ij}$
and $\hat{n}_{\alpha}$ are unit vectors along the $ij^{th}$ bond and in the
$\alpha$ direction.

It is useful to distinguish two contributions to
$\Theta_{ij}$: one due to spin waves, and one due to vortices\cite{nelson81}.
The former is dominant when the $\sin\Theta_{ij} \approx \Theta_{ij}$, while
the other is nonzero when the vorticities $n_{\alpha}(p) \neq 0$.
Thus we write
$\gamma_{\alpha\alpha}= \gamma_{\alpha\alpha}^{SW}+\gamma_{\alpha\alpha}^V$,
where the two terms on the right hand side are respectively the spin wave
and vortex contributions to $\gamma_{\alpha\alpha}$.  The spin-wave
degrees should predominate at low temperatures, while the vortex
degrees of freedom are the dominant excitations near the
phase transition\cite{onsager,feynman55,khoring86,shenoy90,williams}.
  The spin-wave contribution
$\gamma^{SW}$ can be estimated within a self-consistent
harmonic approximation with the result\cite{kleinert89}
\begin{equation}
\label{sqeq}
\gamma^{SW} \sim J \exp \left[ {- k_BT \over 2 D \gamma^{SW}} \right]
\end{equation}
where $D=4$ for a stacked triangular lattice.

\begin{figure}[th]
\begin{center}\leavevmode\psfig{figure=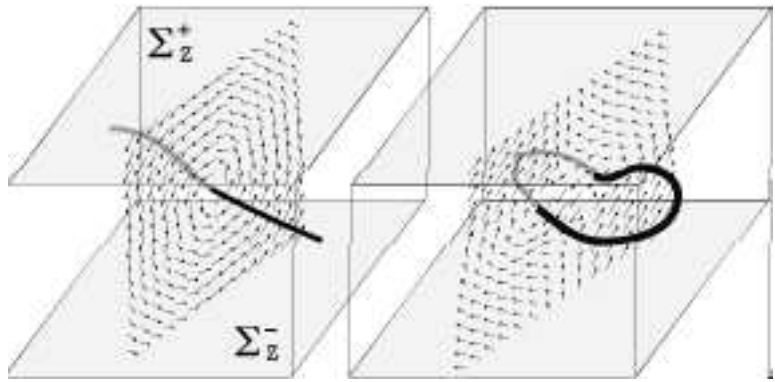,width=4in}\end{center}
\caption [lala] {Two illustrative phase configurations, one with
net vorticity piercing
the sample parallel to the xy-plane (left panel), and one with
zero net vorticity but containing a bound vortex loop
(right panel).}
\label{mdeffig}
\end{figure}

To characterize the vortex
contribution, we introduce the {\em net global vorticity vector} by\cite{chaikin}
\begin{eqnarray}
{\cal M}_\alpha &=& \hat{n}_\alpha \cdot \int_\Sigma d\hat{\sigma} \theta_v
\nonumber \\ &=& \int_{\Sigma^+} \theta_v d\sigma - \int_{\Sigma^-}
\theta_v d\sigma
\end{eqnarray}
where $\Sigma^+$ and $\Sigma^-$ are the two bounding planes normal to
$\hat{n}_\alpha$, with normal vectors parallel or anti-parallel to
$\hat{n}_\alpha$. We assume that the singular portion of the phase variable
$\theta_v$ has been selected out.
${\cal M}_\alpha$ is sensitive to existence of unbound vortex lines 
{\em perpendicular} to $\hat{n}_\alpha$.
This is illustrated in the left panel of Fig.\ \ref{mdeffig} for the case of
a single infinite vortex line piercing the sample normal to the $\alpha$
direction. In this case, the phase integrals on the planes
$\Sigma^+$ and $\Sigma^-$ give nearly equal but opposite values.  Thus
${\cal M}_{\alpha}$ has large fluctuations, leading to a reduction in 
the value of $\gamma_{\alpha\alpha}$ (see below).
Closed vortex loops, such as shown in the right panel,
give a zero contribution to ${\cal M}_\alpha$.
In general, for $f = 0$, the thermal average
$\Big< {\cal M_{\alpha}} \Big> \sim 0$.    For an applied field 
$\parallel \hat{z}$, $\Big< {\cal M}_\alpha \Big> \sim 0$ for 
$\alpha = x$ or $y$.  It can be shown that $\gamma_{\alpha\alpha}$ and
${\cal M}_{\alpha}$ are related by
\begin{equation}
\label{gameq}
\gamma_{\alpha\alpha} \sim  \gamma^{SW}_{\alpha\alpha} -
{1\over V} {J^2 \over k_BT} \Big\{ \Big< {\cal M}^2_\alpha
\Big> - \Big< {\cal M}_\alpha \Big>^2 \Big\}.
\end{equation}
Thus, the vortices make a negative contribution to $\gamma_{\alpha\alpha}$
arising from fluctuations in ${\cal M}$.
They may be said to predominate over the spin waves
when their fractional contribution to the helicity modulus
is of order unity, that is
\begin{equation}
\label{sqcriteq}
{J \over k_BT} \Big\{ \Big< {\cal M}^2_\alpha
\Big> - \Big< {\cal M}_\alpha \Big>^2 \Big\} / {\cal N} \sim {\cal O}(1)
\end{equation}
where ${\cal N}$ is total number of grains.

\begin{figure}[th]
\begin{center}\leavevmode\psfig{figure=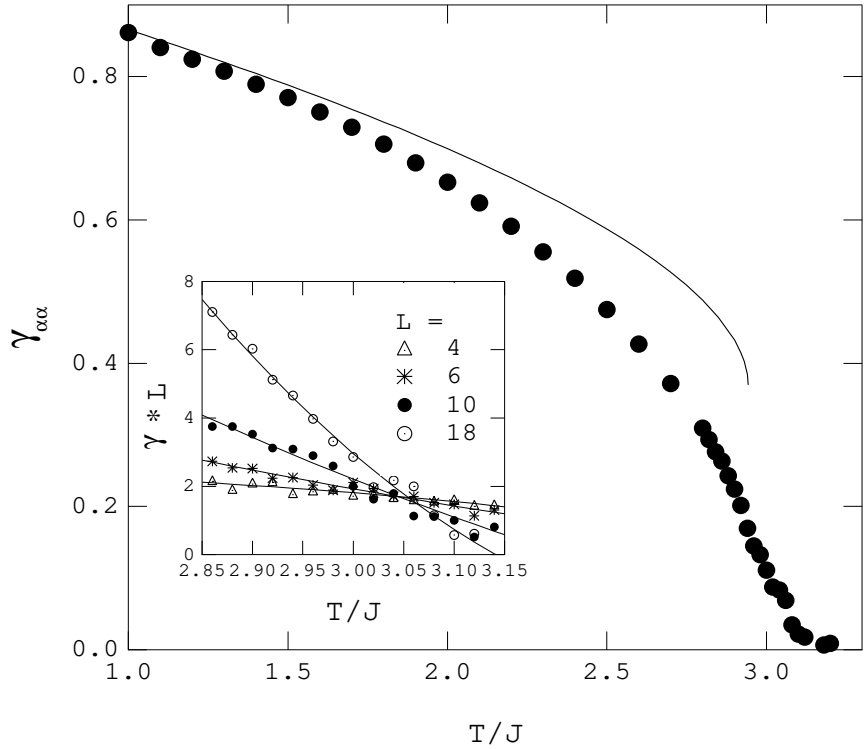,width=5in}\end{center}
\caption [lala] {$\gamma_{zz}$ from Monte Carlo with 10,000 ($T/J<2.7$)
to 50,000 ($T/\ge 2.7$) MC sweeps. The line represents a calculation based
on the harmonic self-consistent approximation. The inset shows the finite size
scaling analysis to locate $T_{XY} = 3.04 \pm 0.02 J$.}
\label{f0gammafig}
\end{figure}

In Fig.\ \ref{f0gammafig}, we show the calculated $\gamma_{zz}$,
as well as the value $\gamma_{zz}^{SW}$ determined from the self-consistent
harmonic approximation (SCHA),  eq.\ (\ref{sqeq}).
The other principal components of $\gamma$ behave similarly.
$\gamma_{zz}$ and $\gamma_{zz}^{SW}$ begin to differ for
temperatures as low as $T \sim 0.3 T_{XY}$,
where vortex loops start to be excited.
The SCHA predicts a discontinuous
jump in $\gamma_{SW}$ from a value of about $0.37$ at
$T_{XY}$ to zero.
This jump is an artifact of the approximation, which neglects
the periodicity of the Hamiltonian in the angle variables and the vortex fluctuations.
The inset shows a finite size scaling analysis to locate $T_{XY}$.
The helicity modulus
$\gamma \sim |T - T_{XY}|^v$ with $v = (d-2)\nu.$
Therefore, the scaled quantity $\gamma L$
for an $L\times L \times L$  system should cross a single point at
$T_{XY}$.  Based on this criterion, our numerical results give
$T_{XY} = 3.04 J \pm 0.02$. We also observe that $\gamma(T)$ approaches
zero with $v \sim 2/3$ for $0.02 < |T_{XY} - T| / T_{XY} < 0.1$ and deviates
from this outside the range.

Eq.~(\ref{gameq}) is equivalent to that derived in Fourier space by 
Chen and Teitel\cite{chen97} for $\lambda \rightarrow \infty$: 
\begin{equation}
\label{chengammaz2}
\gamma_{zz} (q\hat{x}) = J \Big[ 1 - {4\pi^2 J \over V T}
{< n_y(q\hat{x}) n_y(-q\hat{x}) >\over q^2} \Big].
\end{equation}
if we take $q \rightarrow \pi / {\rm L}_x$.

\begin{figure}[th]
\begin{center}\leavevmode\psfig{figure=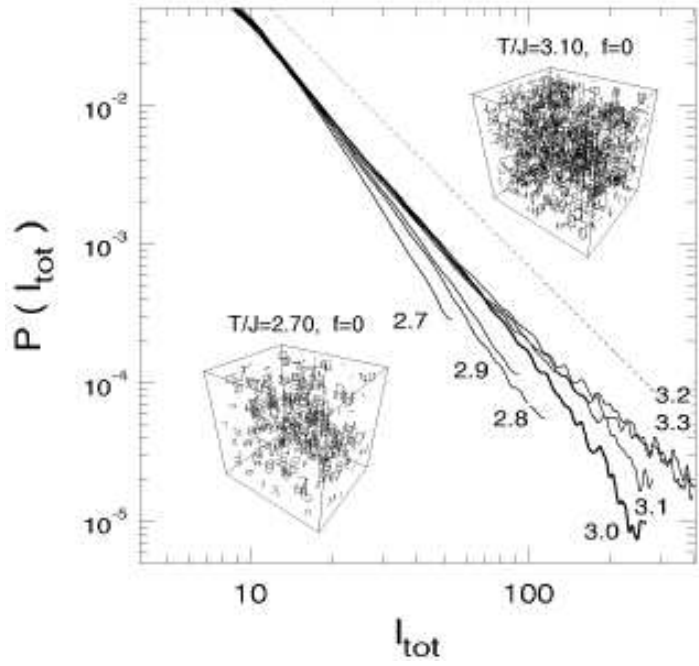,width=4in}\end{center}
\caption [lala] {Size distribution of {\em connected} vortex segments
for $f=0$. The insets show typical vortex configurations for $T/J = 2.7$
and $3.1.$}
\label{f0plfig}
\end{figure}

An alternative form based on the vortex loop scaling picture has been obtained
by Williams\cite{williams95} in the limit
$\lambda \rightarrow \infty$ where $\gamma_{\alpha\alpha}$ is given in terms of vortex 
loop diameter distribution.
To check the importance
of large loops, we show in
Fig.~\ref{f0plfig}, our calculated size
distribution of {\em connected vortex loops} near $T_{XY}$.
Two vortex segments are considered {\em connected} if they cross within
a single unit cell.  Such crossings become very extensive
at $T_{XY}$, suggesting that the energy barrier for
vortex line intersections vanishes near
$T_{XY}$
For $T < T_{XY}$, the maximum vortex loop size
is finite, while for $T \ge T_{XY},$ there start to occur vortex
lines spanning the entire simulation cell.
Thus $T_{XY}$ somewhat resembles a bond-percolation transition, although $T_{XY}$ does
not  correspond to the percolation threshold, but to a point where connected cluster
first forms a 
$D-1$ dimensional manifold if the system is in dimension $D$.
Because such infinite clusters occur, it appears that the behavior seen in
Fig.~\ref{f0plfig} is not a consequence of the finite simulation cell,
but persists in the thermodynamic limit.  A similar
picture, but with no long-range interactions among the vortex segments 
(the so-called polymer limit),  has been discussed by
Akao\cite{akao96} and by Kultanov {\em et al}\cite{kultanov95}.

\subsection{Dissipation near $T_{XY}$}

\begin{figure}[th]
\begin{center}\leavevmode\psfig{figure=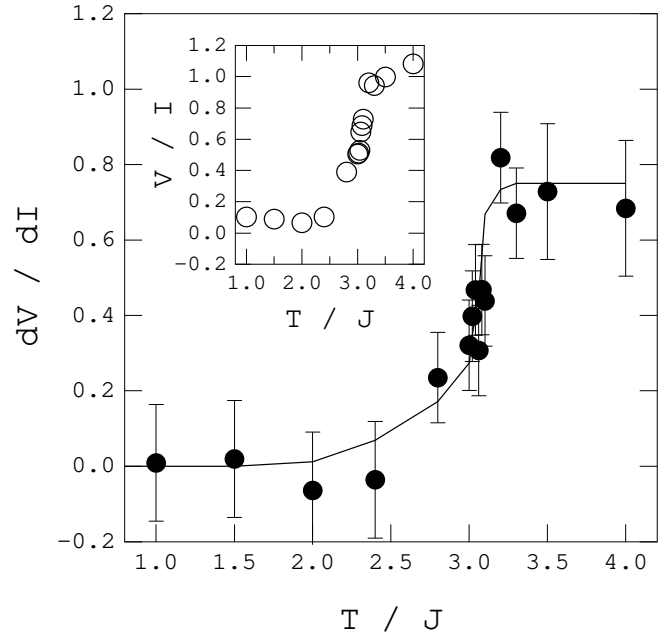,width=4in}\end{center}
\caption [lala] {Dissipation
($dV/dI = [V(I=0.083) - V(I=0.043)] / 0.04$) across the XY transition,
$T_{XY} = 3.04 J.$
A uniform current of $I/I_c$ per grain is flown through yz planes.
The inset shows $\rho \equiv V/I$ at $I= 0.083 I_c$.}
\label{f0rvstfig}
\end{figure}

We have also calculated the dissipation near $T_{XY}(f=0)$, using the periodic
current injection geometry of Fig.\ \ref{geomfig},  In this
case, since there are no field-induced vortex lines, the calculated dissipation
can be unambiguously related to the resistivity.  Our results are shown
in Fig.\ \ref{f0rvstfig} for several values of the bias current density.
[More details of the method are described in section \ref{f24rvstsection}.]
To calculate the differential resistance, $dV/dI,$
we carried out two separate runs at $I/I_c = 0.083$ and $0.043$,
to obtain $dV/dI = V(0.083) - V(0.043)/0.04$.  Fig.\ \ref{f0rvstfig} shows
this result as well as $R \equiv V(0.083) / I$ in the inset.
These bias currents are low enough to show sharp features at $T_{XY}(f=0)$
while not significantly disrupting that transition.
At higher current densities (not shown), there are numerous
current-induced vortex loops.  These increasingly round out the sharp jump
at T$_{XY}(f=0)$ shown in the Figure, which eventually washes away entirely.

\begin{figure}[th]
\begin{center}\leavevmode\psfig{figure=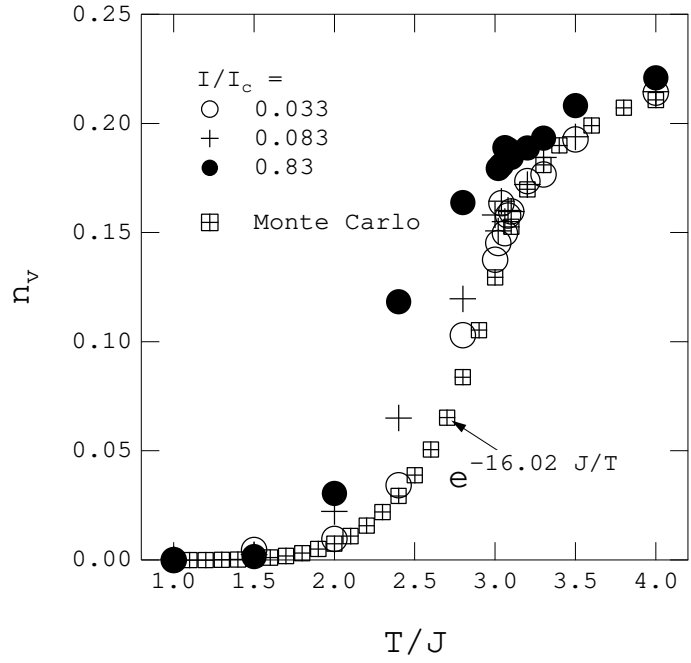,width=4in}\end{center}
\caption [lala] {Average number of vortex segments in equilibrium(Monte Carlo)
and with various bias current densities(RSJJ dynamics). The equilibrium
density
very closely follows an activated form with $U = 16 J$ for $T < T_{XY}.$}
\label{f0nvfig}
\end{figure}

Fig.\ \ref{f0nvfig} shows the average number of vortex segments per plaquette
as calculated both by Monte Carlo simulations (with no driving current) and by
coupled RSJ dynamics (with a finite bias current).
Evidently, just at the XY transition, the system becomes
filled with thermally generated vortex segments, {\em one per
grain, or elemental cell}. Below $T_{XY}(f=0)$, while there may exist vortex
loops of arbitrary size, the number of these, ${\cal P}(l)$, falls off
exponentially for large clusters (cf.\ Fig.\ \ref{f0plfig}).  By contrast,
for $T\ge T_{XY},$ ${\cal P}(l)$ diminishes algebraically with $l$.  This
subtle change in ${\cal P}(l)$ implies that the average size of the
connected vortex tangle  remains finite for $T < T_{XY}$, but diverges above
it\cite{onsager}.  We find that there are numerous {\em finite} vortex loops
at temperatures as low as 0.5$T_{XY}$.  Moreover, at any given temperature,
we find that a large bias current further enhances both their number and
their size.

A simple argument suggests that the dissipation below $T_{XY}(f=0)$ may be
exponentially activated.  The effective energy for a vortex loop
of radius $r$ oriented normal to a uniform driving current density
$j$ is $U(r) \sim 2\pi r \cdot {\rm Min}[ \ln (r) , \ln (\lambda) ]
- c \cdot \pi r^2 j$, where $\lambda$ is the penetration depth.
Thus, there is a barrier to loop expansion with a critical
radius $r_c \sim \ln \lambda  / j$ and height $U_{max} \sim (\ln \lambda )^2 /
j.$ For sufficiently small j, only those vortex loops of size $r > r_c$ will
expand and contribute to dissipation.   From our simulations, for
$T < T_{XY}$, ${\cal P}(l)$ decays exponentially.  Therefore,
the small dissipation involving the expansion of thermally nucleated vortex
rings should have an activated temperature-dependence for
$T < T_{XY}$, resulting in highly nonlinear IV characteristics.

\section{Dense limit: f = 1/6}

This is the largest value we studied which allows a triangular vortex
lattice commensurate with the underlying triangular grid.
This value yields a strong first order
transition\cite{hetzel92,dominguez95},
with an entropy of melting $\Delta S$ of about 0.3 k$_B$ per vortex pancake.

\begin{figure}[th]
\begin{center}\leavevmode\psfig{figure=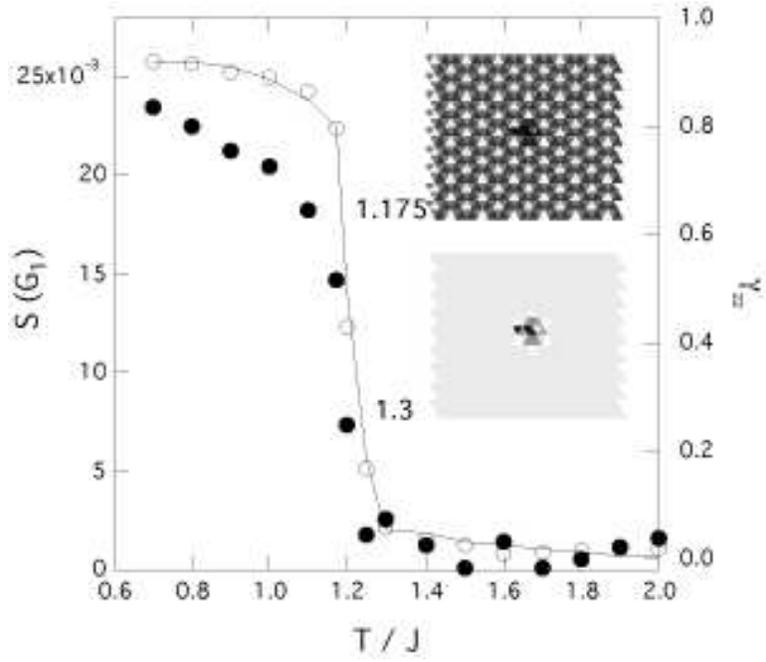,width=4in}\end{center}
\caption [lala] {Normalized Bragg intensity(open circles) and $\gamma_{zz}
$(filled circles) for $f=1/6.$ The solid line is guide for the eyes. The
insets show the real space density correlation $<n_z(r,z) n_z(0,0)>$ for
the local z-vorticity taken over 40,000 MC sweeps.}
\label{f6sqfig}
\end{figure}

In the present MC simulation, the lattice was
gradually warmed up from a perfect triangular lattice in temperature steps
of $dT = 0.1 J$, with $50,000$ MC sweeps for each $T$.
The insets show the in-plane density-density correlation
$< n_z({\bf r}_\perp, z) n_z(0,0) >$ for $T/J = 1.175$ and $1.3$,
slightly below and slightly above the melting temperature.
At the melting temperature $T_m$, our MC histogram for the
internal energy distribution agrees with that of \cite{hetzel92}.
We have also calculated both $\gamma_{zz}(T)$ and
the Bragg intensity $S_{zz}({\bf G}_{1})$ at the smallest
reciprocal lattice vector ${\bf G}_{1}$ for the triangular vortex lattice.
The results are shown in Fig.\ \ref{f6sqfig}.
To within $\delta T/J \sim 0.025$, both
quantities vanish close to $T = T_m(1/6) \sim 1.175 J$,
the melting temperature as determined
from the double peaks in the energy histogram\cite{hetzel92}.
The apparent
melting $T \approx 1.2 J$, slightly higher than inferred from the energy
histogram, seems to be due to a superheating effect.

\begin{figure}[th]
\begin{center}\leavevmode\psfig{figure=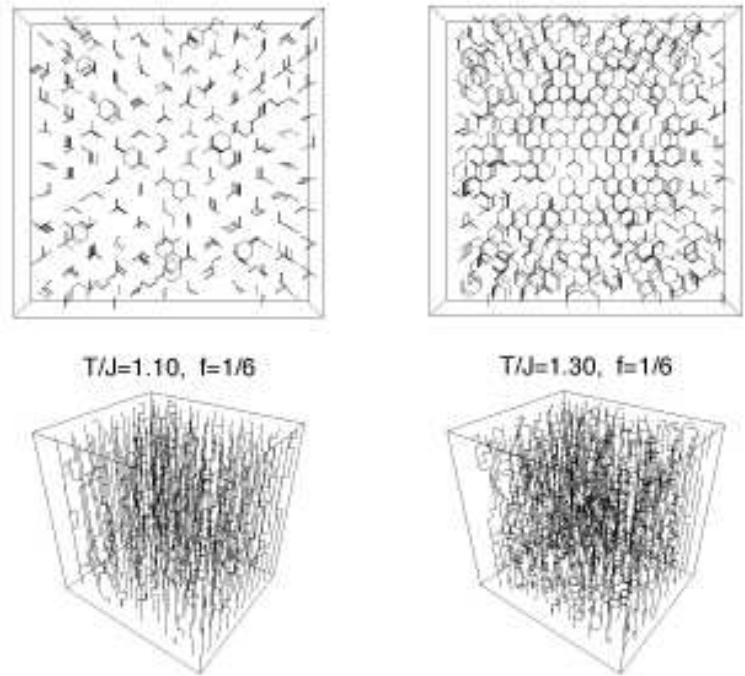,width=4in}\end{center}
\caption [lala] {Typical flux line configurations at for two temperatures
($T/J = 1.1$ and $1.3$) spanning the melting temperature at $f=1/6$,
plotted for an $18\times 18\times 18$ grid.
The upper panels are top views.}
\label{f6linesfig}
\end{figure}

The occurrence of only a single phase transition at
$f = 1/6$\cite{hetzel92,jagla96} is not surprising:
at this field, there is one vortex
pancake for every three grains, and hence, only
three phase degrees of freedom per pancake.
Thus, most potential vortex excitations are
already exhausted by the lateral fluctuations of field-induced
vortex lines in the liquid phase.  This can be seen in Fig.\
\ref{f6linesfig}, where we show two typical vortex configurations,
one slightly below and the other above the melting transition.
Clearly, the transverse line fluctuations quickly dominate the
thermodynamics above $T_m$.
The strong first order transition at $f=1/6$ can be then
understood by the close connection between the lateral line fluctuations
and incipient vortex loops.
We conclude, with an accuracy of $dT/J = 0.025$,
that in the dense limit of $f = 1/6$, superconducting coherence is destroyed
in all directions as soon as the lattice melts.

\section{Dilute limit}

By {\em dilute limit}, we mean the regime where the number density of thermally
excited vortex line segments $n_c$ equals or exceed the density of
field-induced vortex
line segments.  To make this more quantitative, we first
consider a perfect line lattice at $T=0$ with a given $f$. The number of
unit vortex line segments {\em per grain} will be
$1/(3f)$, or $1/(15f)$ per plaquette (since there are five plaquettes per
grain).  Thus, for $f=1/6$, there are 0.4 vortex segments per plaquette.
In Fig.\ \ref{f0nvfig}, we show the density n$_c$ of {\em thermally excited}
vortex segments at $f = 0$, including all three directions.
Note that at the XY transition, $n_c \equiv 0.15$.
Thus, $f=1/6$ is clearly in the dense regime, while
$f < 1/18$ is roughly in the dilute regime.

\begin{figure}[th]
\begin{center}\leavevmode\psfig{figure=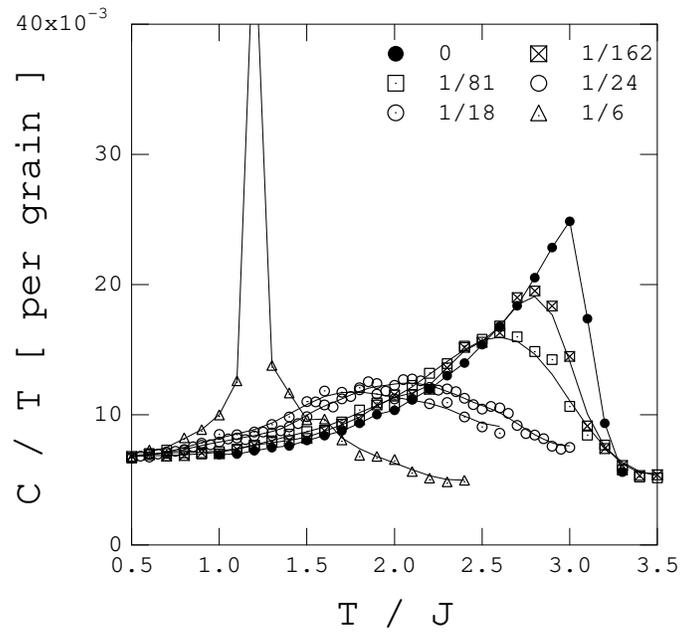,width=4in}\end{center}
\caption [lala] {Specific heat per grain for various frustrations.
For $f=1/6$, $C_v$
at $T = 1.2$ shows a clear divergent behavior, apart from other dilute
cases.}
\label{shfig}
\end{figure}

Fig.\ \ref{shfig} shows the specific heat $C_V$ per grain, as calculated from
energy fluctuations for several
values of f [1/162 (4 flux lines), 1/81, 1/24, 1/18 and 1/6 (108 lines)] in
both the dilute and dense regime.  At low $T$, all the $C_V$'s
approach $k_B/2$ per grain, as expected from the Dulong-Petit law.  The
overall behavior of the peaks in $C_V$ up to $f = 1/24$ is remarkably similar
to that seen in \ybco
\cite{overend94}.   For $f = 0$, it is known that $C_V$ has a weak divergence
($\propto |t|^{-\alpha}$ with $\alpha \approx 0.0$\cite{gottlob93}), as
expected for the $3D$ $XY$ model.
At finite $f$, this peak is rounded as
seen experimentally\cite{overend94,roulin96}.  As discussed below,
these broad peaks in $C_V$ generally occur {\em well above} the melting transition in
this field range.  Note that
our results for the dense ($f=1/6$) case differ qualitatively from
all those at lower $f$.
The sharp peak for $f=1/6$ is actually a delta-function singularity,
consistent with the finite heat of fusion of a first-order transition
known to occur at this density\cite{hetzel92}.

\begin{figure}[th]
\begin{center}\leavevmode\psfig{figure=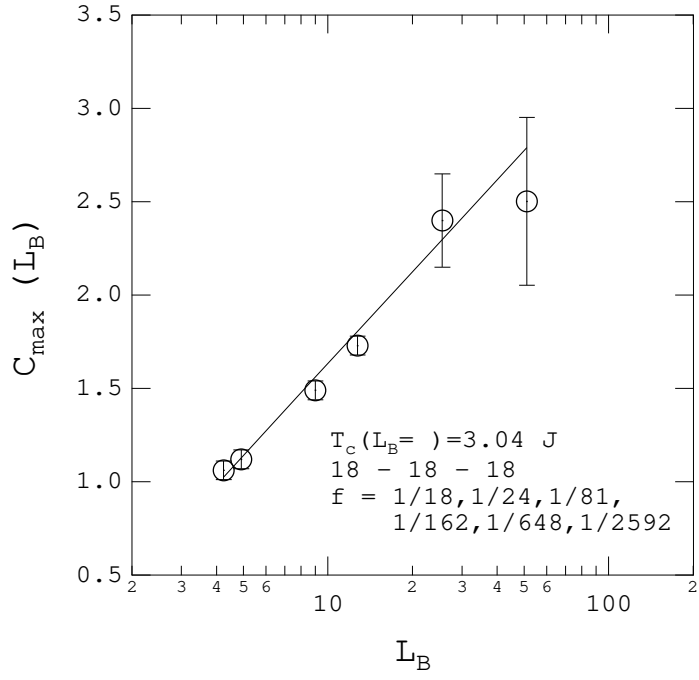,width=4in}\end{center}
\caption [lala] {Dependence of peak height of $C_v$ on the magnetic
length $L_B$. The line is guide for the eyes.}
\label{cpeakfig}
\end{figure}

\begin{figure}[th]
\begin{center}\leavevmode\psfig{figure=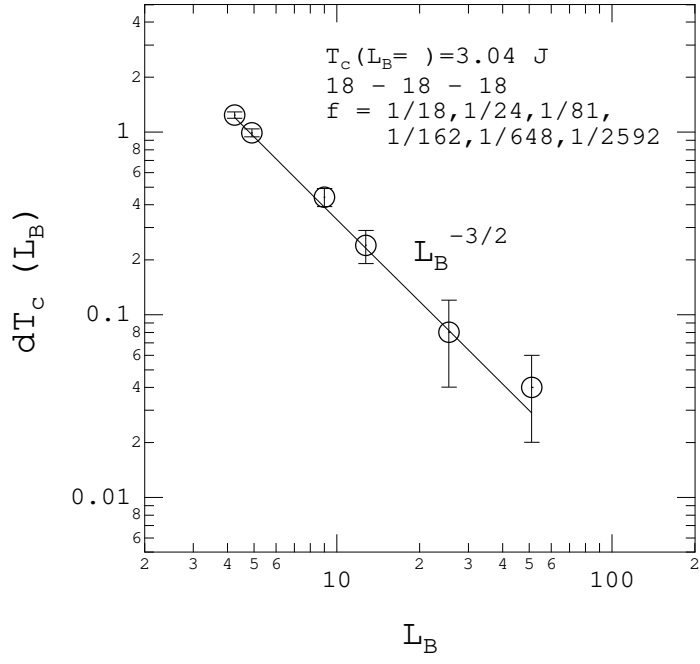,width=4in}\end{center}
\caption [lala] {Shift of peak in $C_v$ from the $f=0$
position($T^0_{xy}$) vs.~the magnetic length $L_B$. The line($\sim
L_B^{-3/2}$) is guide for the eyes.}
\label{deltatcfig}
\end{figure}

Fig.\ \ref{cpeakfig} shows that the height of the peak for
$f \le 1/18$  roughly follows a logarithmic dependence on the magnetic length
$L_B$  defined as $L_B = 1 / \sqrt{f},$ the average vortex spacing.
Furthermore, as we show in Fig.\ \ref{deltatcfig}, the position of the peak
at a finite $f$ shifts from $T_{XY}(f=0)$ by an amount $\delta T_c (L_B)$
which closely follows the law $\sim L_B^{-3/2}.$
As we will show in more detail for $f=1/24$ below the phase rigidity along the
applied field, as measured by $\gamma_{zz}$, vanishes for all $f$ near the
broad maximum in specific heat.
At the temperatures [which we denote $T_\ell(f)$] where
$\gamma_{zz}(f)$ vanishes, we observe that
the average number of thermally generated vortex segments {\em per grain}
normal to the $\hat{z}$ direction closely follows the law
$n^c_{xy}(T_\ell) \sim 0.1 \cdot L_B^{0.6 \pm 0.1}$.  All this behavior
is discussed in more detail below.

\section{f=1/24}

\subsection{Statics: Melting and $\gamma_{zz}$}

In the dilute regime, such as $f \leq 1/24$,
there are numerous phase degrees of freedom per field-induced
vortex pancake.  Thus, a double transition, if there is one, 
might be more plausible here than at $f = 1/6$, one transition
being the melting of the field induced flux lattice, the other
connected to the XY-degrees of freedom\cite{tesanovic95}.

\begin{figure}[th]
\begin{center}\leavevmode\psfig{figure=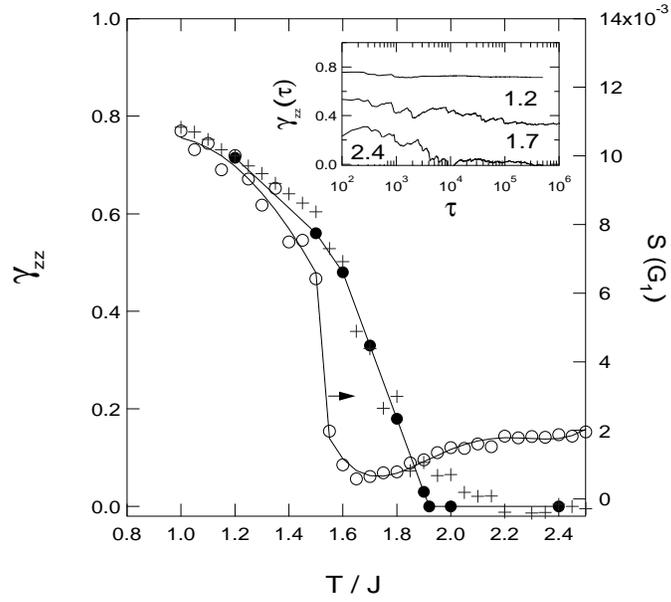,width=4in}\end{center}
\caption [lala] {Normalized Bragg intensity (open circles;
$5\times 10^4$ MC sweeps)
and $\gamma_{zz}$ (filled circles: $5\times 10^5-10^6 $ MC sweeps; crosses:
$5\times 10^4$ MC sweeps)
for $f=1/24.$ The solid lines are  guides for the eyes. The inset shows the
dependence of $<\gamma_{zz}>$ on the accumulation time,
as discussed in the text. }
\label{sq_g_24fig}
\end{figure}

To check this possibility, we have studied
$f = 1/24$ (a field which allows for a commensurate triangular flux
lattice of 48 lines) on a stacked triangular grid of $24\times 24 \times 24$
grains.   We first did an extensive simulated annealing run on a single layer,
verifying that the vortices freeze into a perfect triangular lattice.
We then stacked 24 such layers
to form a three dimensional ground state. Next, the lattice was gradually
warmed up in intervals of $dT/J = 0.1$ or $0.05$, typically with 50,000
MC steps for each temperature.
For several $T$ close to a transition, we ran
up to $10^6$ MC steps to ensure equilibration.
The resulting Bragg intensity
$S({\bf G}_{1})$ and helicity modulus component $\gamma_{zz}(T)$ are
plotted in Fig.\ \ref{sq_g_24fig}.
[The transverse components $\gamma_{xx}(T)$ and
$\gamma_{yy}(T)$ fluctuate around zero for most $T > 0$,
as expected for a very weakly pinned vortex lattice which is
free to slide in the $ab$ plane.]

The results do indeed suggest the possibility of {\em two} phase
transitions.  The first - the melting of the vortex lattice - occurs near
$T = 1.5 J \equiv T_m$, where the Bragg intensity drops sharply.
At higher temperatures, there is a broad dip in the normalized Bragg
intensity which reaches a plateau at around $T/J \sim 2.1.$
The possible upper transition, near $T = 2.0J \equiv T_{\ell}$,
is the point where $\gamma_{zz}(T)$ vanishes.
Essentially the same behavior, but with an even wider temperature separation, 
has previously been
observed on a cubic grid by Li and Teitel\cite{li93_1}.

We have carried out several checks to see if
the separation of these two transitions is an artifact due to a finite-size
effect.  First, as shown in the inset,
we monitored the dependence of $\gamma_{zz}$ on accumulation time $\tau$ up to
$10^6$ MC steps. More precisely, we define $\Big< A \Big>_\tau \equiv
{1\over \tau} \int_0^\tau  A(t) dt$ and
use this in calculating the averages which define $\gamma_{zz}$
in eq.\ (\ref{heldefeq}).
The $<\gamma_{zz}(T)>_\tau$ thus defined generally evolves approximately
logarithmically in $\tau$\cite{logt} until it reaches its apparent
equilibrium value.  For temperatures $T_m < T < T_\ell$,
the system tends very slowly towards an apparently {\em finite} limiting
value.   We have also checked the size dependence up to
$24\times 24 \times 48$, verifying that the $24 \times 24\times 24$ behavior
represents the asymptotic limit.
Li and Teitel have carried out similar checks up to $200$ layers
in the cubic model\cite{li93_2}.  Nonetheless, $\gamma_{zz}$ has  some size-
dependence to a degree dependent strongly on anisotropy of the system. 

If the ratio $J_z/J_{xy}$ is increased to 4.0 (where $J_z$ and $J_{xy}$ are
the couplings perpendicular and parallel to the triangular plane),
the separation $(T_{\ell} - T_m)/J_{xy}$ between the melting transition
and the upper possible transition actually grows for a given size.
For these values, the smallest-${\bf Q}$ Bragg peak
vanishes at $T_{m} \sim  2.9 J_{xy}$, while $\gamma_{zz}$ vanishes at
$T_{\ell} \sim 4.0 J_{xy}$.
On the other hand, for weakly coupled layers with $J_z/J_{xy}=0.1$,
the two transitions merge to within
less than $0.1 J_{xy}$, as in the
the {\em isotropic} dense $f=1/6$ case.
In this weakly coupled case, $T_{m} \sim 0.6 J_{xy}$.

These observations suggest that
phase coherence at finite $f$
in a disorder-free system may possibly be destroyed in two steps.
First, coherence transverse to the average field direction is lost through
melting of the lattice.  But longitudinal coherence persists
until it is destroyed, along with
line-like correlations of the individual vortex segments,
at a slightly higher temperature $T_\ell$.  
This is most apparent for the {\em isotropic} system only when $f \le 1/18.$

\subsection{Vortex Analysis of Possible Transition at $T_\ell$}
\label{vortexsection}
\begin{figure}[th]
\begin{center}\leavevmode\psfig{figure=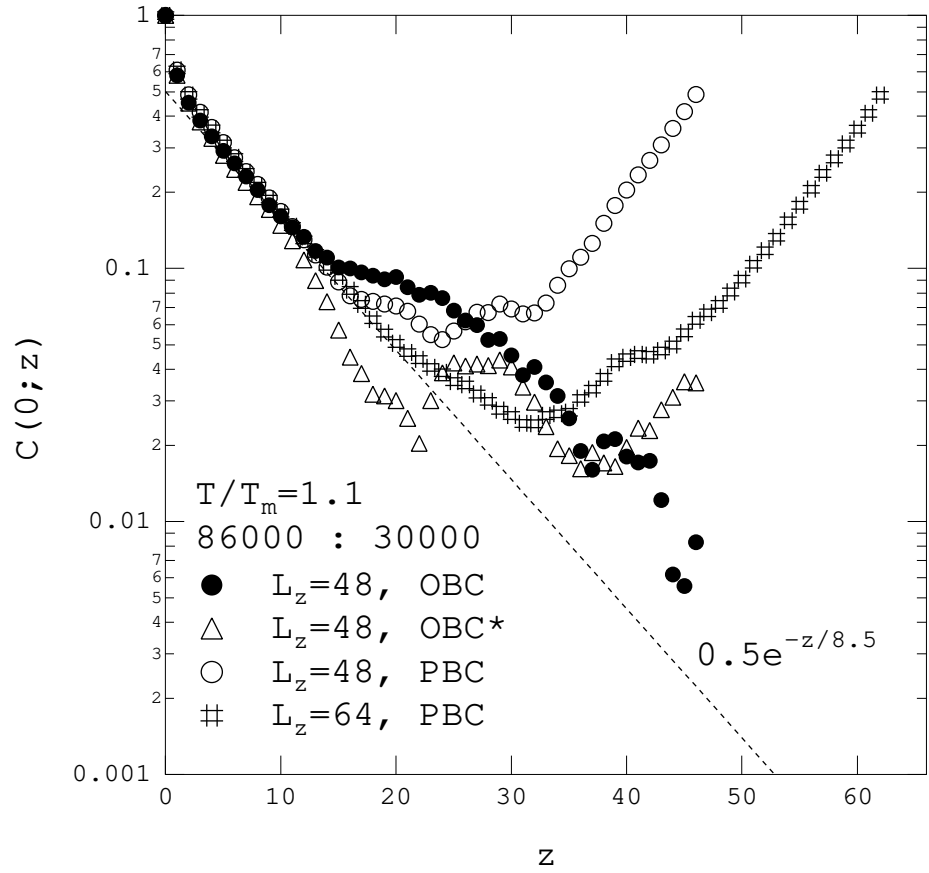,width=5.2in}\end{center}
\caption[lalal]{Phase correlation function $c({\bf 0};z)$ for various
system sizes
and boundary conditions.}
\label{czfig}
\end{figure}

There are several ways to look at general phase correlation function
$<\Theta({\bf \rho}, z) \Theta({\bf 0}, 0) >$ where $\Theta$ is the
gauge-invariant
local phase\cite{glazman91} of the superconducting order parameter.
To probe the
longitudinal phase
coherence, we only consider $c(0;z) \equiv {1 \over A}
\int d^2\rho
<\exp [ i (\Theta({\bf \rho}, z) -
\Theta({\bf \rho},0) ) ] >
$, where $A$ is the sample area - that is, 
the correlation function in the $\hat{z}$ direction. Glazman
and Koshelev
have pointed out that phonon-like fluctuations in the vortex lattice
lead to a
power law decay of $c(0;z)$\cite{glazman91} (not explicitly shown in the
Figure \ref{czfig}).  We observe that this holds
true for
$T < T_m.$  For higher temperatures, this dependence changes to an 
exponential decay
$c(0, z) \propto
\exp (- z / \xi_{0z})$.  The 
correlation length $\xi_{0z} \sim [ \log ({T^2\over T_{0}^2} {\log
(T/T_0)  \over
2\pi^2}) ]^{-1}$
 where $T_0$ is the temperature scale such that $<[\Theta(0,z) -
\Theta(0,z+1)]^2> \sim 1$. 
This behavior is shown for
$T/T_m = 1.1 $ in Fig.\
\ref{czfig} for systems of several thicknesses. To ensure equilibration, we
ran 86000 MC
sweeps before accumulating data over the following 30000 MC sweeps. To
check the effect of
boundary conditions, we used both open boundary (OBC) and periodic boundary
conditions (PBC)
along $\hat{z}$; periodic boundary condition was used in the xy plane for
both cases.
In all cases, we observe a robust exponential dependence over a limited range $1
< dz < \xi_x$
where deviation sets in at $dz \sim \xi_x \sim 12 $ for $T/T_m = 1.1$, for example.
In this temperature range,
we do not find significant dependence of c(0, z) on either system size 
or boundary conditions.
For a given temperature above $T_m$, we can use this robust temperature
regime to
extract the
phase correlation length $\xi_{0z}$ from our numerical simulation.

\begin{figure}[th]
\begin{center}\leavevmode\psfig{figure=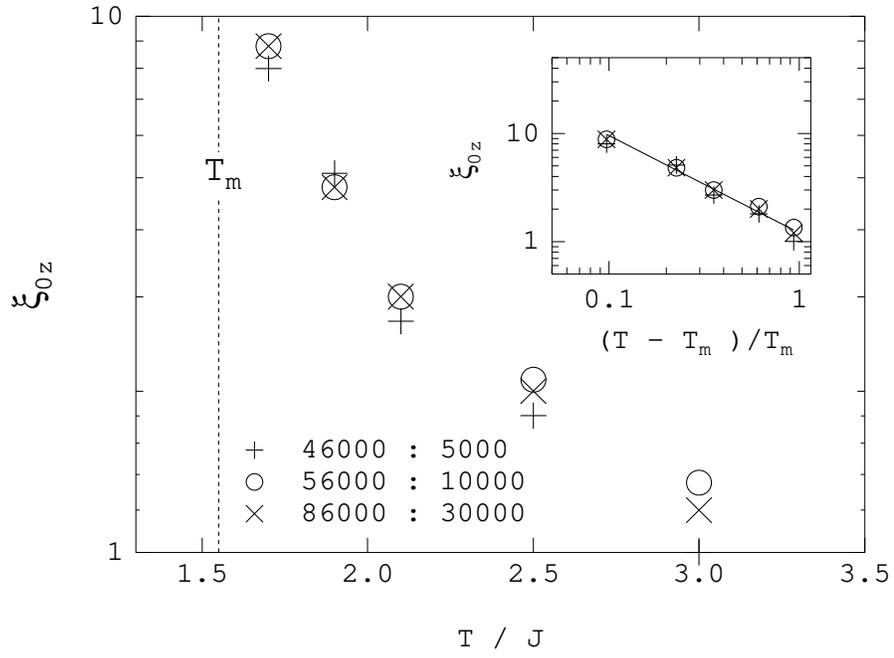,width=5in}\end{center}
\caption[lalal]{Longitudinal phase correlation length $\xi_{z0}$ determined
from
fitting $c({\bf 0};z)$ to the form $a \exp [-z / xi_{z0}]$ for $z <
\xi_x(T)$ as
discussed in the text.}
\label{xizfig}
\end{figure}

The result is shown in Fig.\ \ref{xizfig}.
For all $T > T_m$, and for separations less than $\xi_x(T)$, 
we observe that $c(0, z) \sim \exp[ - z / \xi_{z0} ]$. 
$\xi_{z0}$ gradually decreases with increasing temperature approximately
as $(T-T_m)^{-1}$, becoming equal to the unit layer spacing near
$T/J = 3.0 \sim T_{XY}(0)$ as shown in the inset. 
We also observe(not shown in the Figure) 
that $\int d{\bf \rho} c({\bf \rho}, z)$ has
far milder dependence on $z$.
The exponential decay in $c(0, z)$ is accounted for by random walk-like excursions
of the vortex lines and the presence of dislocation or disclination 
loops in this temperature regime. Note that spinwave excitations in the 
vortex lattice usually lead to an algebraic decay of $c(0, z)$.
Note that given a {\em static} deformed vortex line configuration, we may still find a 
coordinate transformation $\{x, y, z\}\rightarrow \{x',y',z\}$ into a curved space in
which the vortex line is straight.  In that coordinate system, we will have a long
range phase coherence along the straight line in the $\hat{z}$ direction. 
Therefore, the apparent exponential decay of $c(0, z)$ is not an equivocal indicator
for the  destruction of phase coherence along $\hat{z}$, but gives information about
the deformation of vortex lines from the straight configuration. 

In the large-distance tail of $c(0;z)$, where $z > {\rm L}_z/2$,
$c(0, z)$ does depend on the boundary conditions and system size, having an upturn for the 
periodic boundary condition as expected.
Moreover, $\xi_{z0}(T)$ rapidly falls as $T$ decreases 
toward $T_{XY}$ from above, leaving a large interval $\xi_x(T) < dz  \ll {\rm L}_z/2$ in which 
the behavior deviates from simple exponential decay and is independent of the boundary condition 
used. We also observe that the deviations have relatively poorer statistics due to slow
kinetics.  It is likely to originate from 
vortex entanglement and cutting/reconnection, which develop on length
scales larger than
$\xi_{z0}$.  Each of these rare and slow vortex
crossings produces
a drastic and long-lasting impact on the local phase correlations. The
rarity of these events is due to the sizable barrier for vortex cutting and
to the subdiffusive
nature of vortex lines motion.  
Because of this rarity, a rapid thermal
cycling across this sluggish region may lead to hysteresis.

\begin{figure}[th]
\begin{center}\leavevmode\psfig{figure=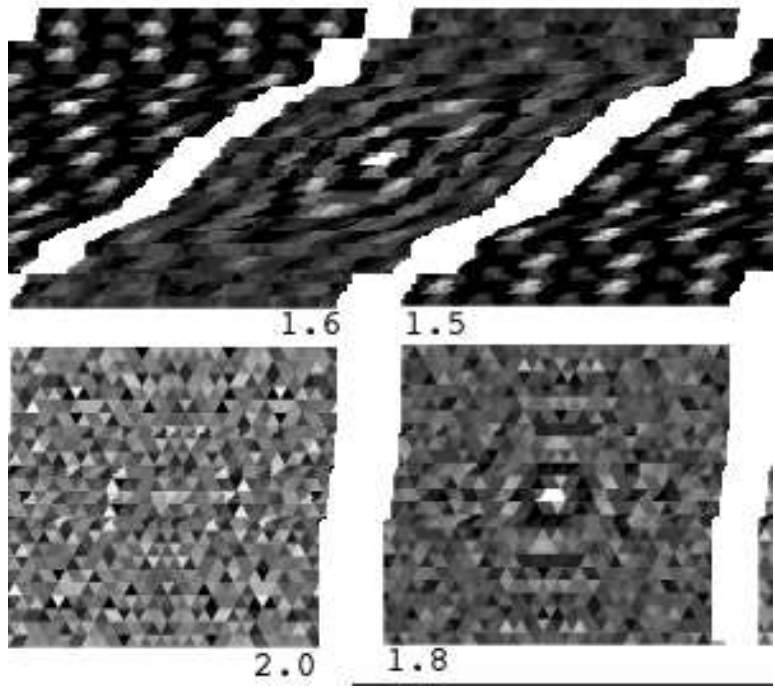,width=4in}\end{center}
\caption [lala] {Real space top-to-bottom density correlation function
$<n_z(r,z=12) n_z(0,0)>$ for $T/J = 1.5,
1.6, 1.8$ and $2.0$  for $f=1/24$ in a $24\times 24 \times 24$ grid.
Melting occurs around $T_m/J = 1.55$ while
line-like correlations (white spot at the center) vanish around
$T/J = 2.0 \sim T_{\ell}/J.$  50,000 MC steps for each temperature.
}
\label{cor24fig}
\end{figure}

We now look more closely into
the vortex configurations near $T_\ell$ for $f=1/24.$
Fig.\ \ref{cor24fig} shows the density-density correlation function
$n_{2,z}({\bf r}_{\perp}, {\rm L}_z/2) = < n_z({\bf r}_{\perp},{\rm L}_z/2)n_z(0,0)>$
describing the $\hat{z}$-component of local vorticity at separations equal to
half the total thickness ${\rm L}_z$ (=24). The most prominent feature in $n_{2,z}$
is the
disappearance of  triangular correlations in the $xy$ plane at melting ($T_m
\sim 1.5 J).$  This behavior is consistent with the disappearance of the Bragg
spots in Fig.\ \ref{sq_g_24fig}.  Note, however, that the
central spot, corresponding to the self-correlation between the two ends
of the same line, persists well
above melting until it vanishes near $T/J = 2.0$, close to $T_\ell.$
This is consistent with the fact that line-like correlations are maintained
over at least 12 layers up to $T/J = 2.0$ as we 
already noted in Fig.\ \ref{xizfig}.

In a sample
of truly macroscopic thickness, the lines in the liquid phase should
carry out random-walk-like excursions, leading to loss of
top-to-bottom vortex density correlations over a finite correlation length
denoted $\xi_{vz}(T)$, which may be of the same order of magnitude as $\xi_{0z}$ defined 
above.
A finite $\xi_{vz}(T)$ means that
the underlying lines are {\em flexible}, not that they break up
into 2D vortices.  This breakup becomes relevant only for $T>T_\ell$.
The objects which break apart into 2D objects
above the melting transition are not the lines themselves, but the
{\em topological defects of the lattice}, such as disclinations, which tend to
appear as {\em well-aligned} 
line defects near $T_m$\cite{ryu96,kim96}. Topological defects look well aligned only when 
$|{\bf r}_{i}(z) - {\bf r}_{i}(z+1)| / a_B
\ll 1$  where ${\bf r}_{i}(z)$ is the position of the segment of vortex line $i$ in the $z$-th layer.
Note that the relevant minimum length scale for alignment of defect lines 
is the mean vortex spacing, $a_B$.  
The destruction of the {\em lattice order} along the field, which may be detected by 
vanishing Bragg peak in neutron diffraction, 
is related to proliferation and unbinding of these defects.  
On the other hand, the destruction of phase 
coherence along ${\bf B}$, as we discuss in more detail later, is related to the presence of 
fluctuations in the transverse vorticity. As the vortex density decreases, it is not {\em a
priori} obvious if the energy scales for these different types of defects should remain the same.

Thus the line-liquid regime, if it is really a distinct thermodynamic phase, 
may possibly be described
as a neutral gas of topological defects (disclinations of both signs) within the triangular
lattice in each plane, which are
correlated over a finite length in the $\hat{z}$ direction.
Above melting, one expects unbound disclinations
to proliferate.  Hence, the long-range structural correlations
of the vortex lattice are lost in all directions upon melting.
However, the {\em phase rigidity}, as measured
by $\gamma_{zz}$, may persist even
above melting, but scaled down by the factor of
$\xi_{z0}/{\rm L}_z$, 
the fraction of the volume of the sample into which the
applied twist penetrates.
Presumably, this continuous suppression, unless pre-empted by a first order
transition (for high densities),
persists until the condition ${\rm L}_z/\xi_{z0} \rightarrow \infty$ is met 
via proliferation of  ``unbound vortex loops'' (vortex lines extending an
infinite distance in the transverse direction). At this point, the phase coherence even between 
neighbouring planes normal to ${\bf B}$ will be lost.

\begin{figure}[th]
\begin{center}\leavevmode\psfig{figure=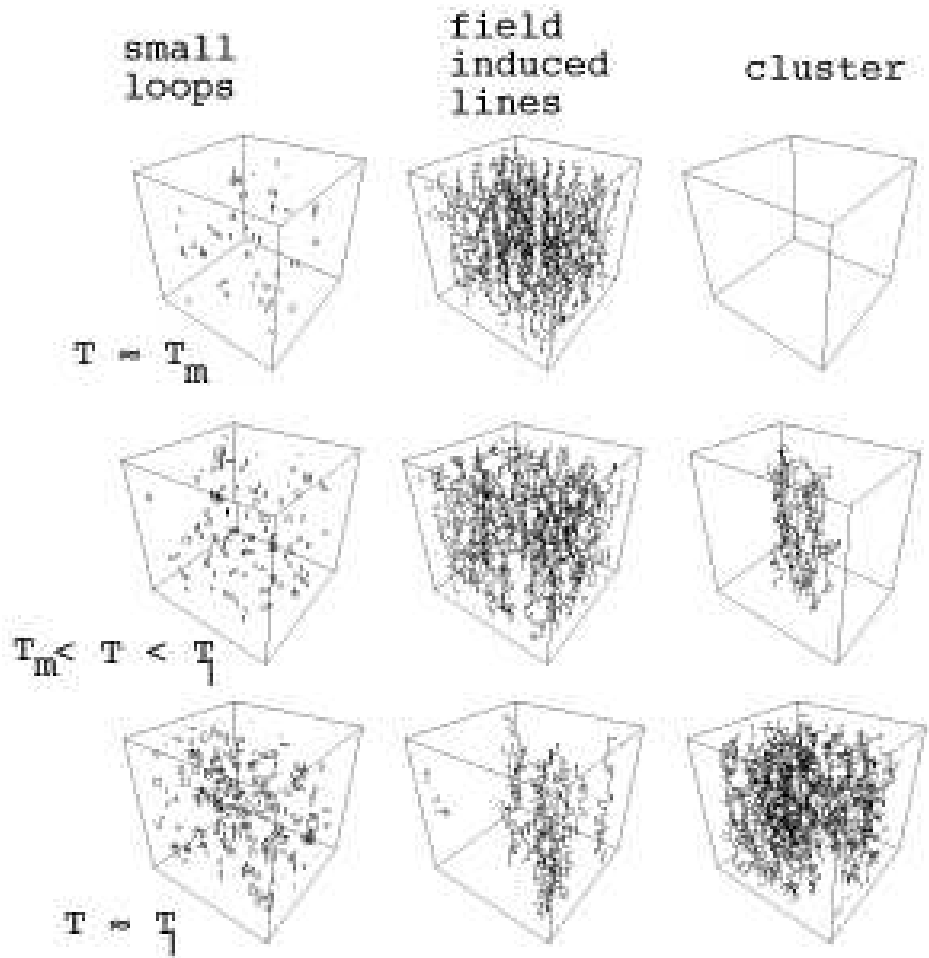,width=5in}\end{center}
\caption [lala] {Vortex configurations for temperatures $T$ in the range
$T_m \le T \le T_{\ell}.$
The left column shows the portion of vortex fluctuations which form
bound loops. The center column shows mainly those field-induced vortex lines
which are not entangled, while the right column shows the largest cluster
of entangled lines (as defined in the text).
}
\label{f24clusterfig}
\end{figure}

In Fig.\ \ref{f24clusterfig}, we show snapshots of vortex configurations at
T$_m$, T$_{\ell}$ and a temperature between $T_m$ and $T_{\ell}$.
In this regime, by using a bond-searching algorithm,
we have identified three distinct classes
of vortex lines.  The first consists of small vortex loops
which close on themselves
without crossing either of the two opposite bounding surfaces.
The second class contains all isolated
lines beginning at the bottom $xy$ plane and ending at the top one.
Most of the
disentangled field-induced vortex lines fall into this group.
Finally, there occur ``vortex tangles''.  These are lines
connected at a given time to one another by the crossing of two vortex
segments in the same unit cell.
Such tangles are formed either by collision of two flux lines or by
interactions of such lines with the vortex loop excitations.
This tangle  is not static:
the collisions which produce it are more and more frequent 
with increasing temperature and its overall shape will evolve with more rapidity as $T$ 
increases.

The three columns of the Figure represent the fraction of the
vortices belonging to each class at a given instant.
On melting ($T/J = 1.5$), the fluctuating lines in our finite
sample still remain largely disentangled and separated from each other.
As  $T \rightarrow T_\ell,$  the density
of loop excitations increases (left column), while
the field-induced lines (central column) have stronger
lateral fluctuations .
Both of these effects cause more and more ``connected" clusters (i. e., vortex
tangles) to appear.
Finally, at $T_\ell,$ an {\em infinite} tangle, connected by
crossing vortex lines, forms.
At this temperature, the connected tangle of vortices form a $(D-1)$ dimensional 
manifold of a tortuous shape, transverse to ${\bf B}$, and cut the original $D(=3)$-
dimensional coherent XY system into halves.

\begin{figure}[th]
\begin{center}\leavevmode\psfig{figure=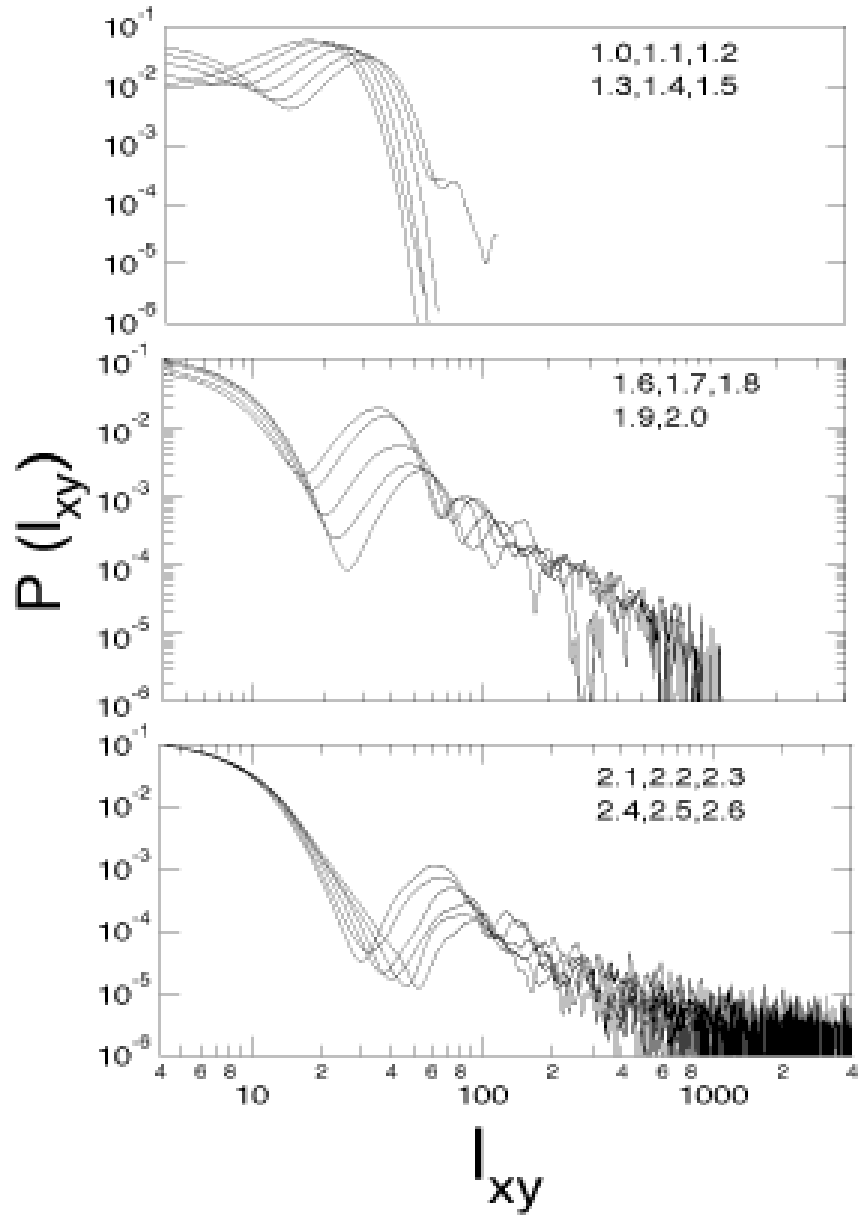,width=5in}\end{center}
\caption [lala] {The distribution of transverse vortex lengths $\ell_{xy}$
projected onto the xy-plane $\ell_{xy}$ for three sets of temperatures in the
(a) lattice, (b) line liquid, and (c) tangled vortex web states.
}
\label{f24plfig}
\end{figure}

Fig.\ \ref{f24plfig} shows
an instantaneous vortex cluster size distribution for
various temperatures at $f = 1/24$.  To generate this distribution,
we define the {\em projected transverse length} of each vortex loop
(or tangle) by $\ell_{xy} \equiv \oint | \hat{z}\times {\bf n}_v | $
for each isolated cluster composed of unit vortex segments ${\bf n}_v$,
and accumulate a histogram, ${\cal P}(\ell_{xy})$.  
[We consider
only the size distribution projected onto the xy plane because
the field induced lines (which are infinite along $\hat{z}$)
could mask the loops with large extent in the z direction.  We
also believe that these fluctuations are more relevant to the
vanishing of $\gamma_{zz}$.]

In the first panel of Fig.\ \ref{f24plfig}, we plot ${\cal P}(\ell_{xy})$
for several $T \le T_m$.  Each plot has a sharp maximum cutoff and
a pronounced peak, which is due to the finite
average lateral fluctuations of the field-induced vortex lines.
For $T_m < T < T_\ell$ (second panel), the weight of distribution is shifted towards
larger $\ell_{xy}$, because lines in the liquid phase undergo larger transverse
fluctuations.   Closed vortex loops also begin to appear in this region.
As $T$ increases, the distribution is cut off at progressively larger values,
as more and more lines join the connected clusters.  Finally,
for $T> T_\ell$, all curves are
characterized by the  appearance of ``infinite" clusters, with no obvious
length cutoff.   The distribution appears to fall off algebraically
in this regime - i.\ e., ${\cal P}(\ell_{xy}) \sim
\ell_{xy}^{-\mu}$ with $\mu \sim 1.0 < 2$ - suggesting $<\ell_{xy}>\rightarrow
\infty$ for $T > T_{\ell}$.

\begin{figure}[th]
\begin{center}\leavevmode\psfig{figure=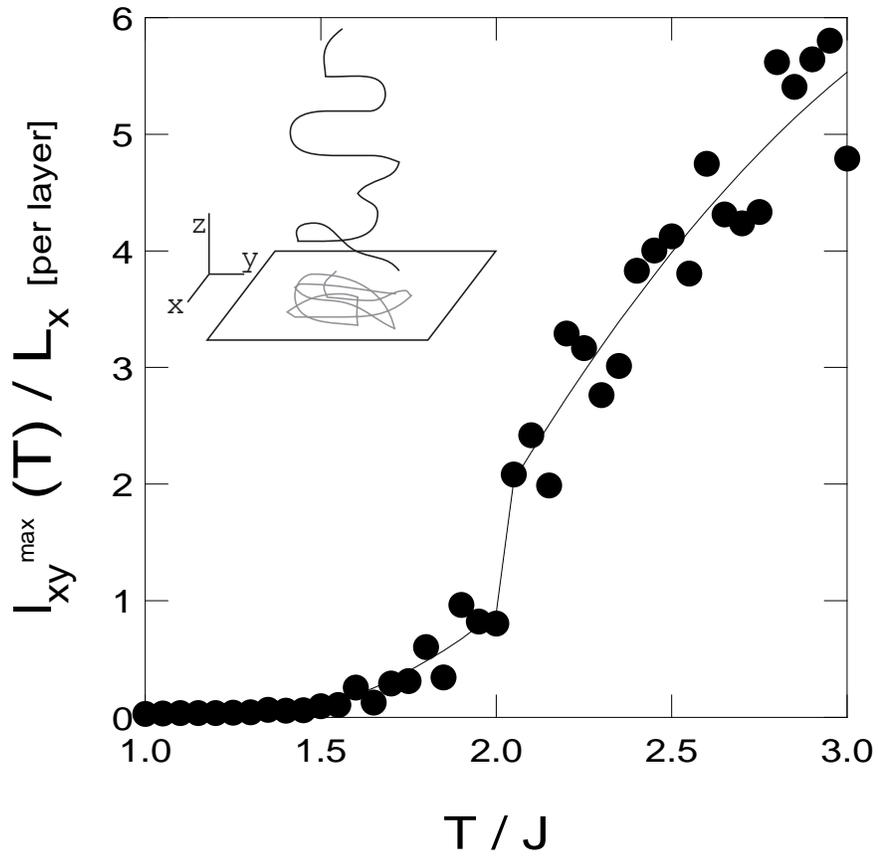,width=5in}\end{center}
\caption [lala] {Maximum projected lateral vortex length $\ell_{xy}$,
given in terms of the lateral box dimension (=48), and
normalized per layer in a $24\times 24\times 24$ grid
with $f=1/24.$}
\label{f24plmaxfig}
\end{figure}

In Fig.\ \ref{f24plmaxfig}, we show the maximum
value $\ell_{xy}$ occurring over
$10,000$ MC sweeps for each temperature.
The size is normalized by the linear system dimension in the $xy$ plane
(48), and also by the number of xy-planes (24).
Although $\ell_{xy}$ grows monotonically for $T>T_m,$
it seems to jump discontinuously from $\sim 1$ to 
$\sim 2$ between $T/J = 2.0$ and 2.1 (near $T_\ell$ for samples of this size).
Qualitatively similar behavior occurs for the isotropic
$f=0$ XY transition near $T_{XY}(f=0)$
[cf. Fig.\ \ref{f0plfig}].

\subsection{Entanglement, Winding Number and Other Exotica}

We now discuss 
a possible extension of the vortex loop picture of the zero field XY
transition to the hypothetical $T_\ell$ transition or crossover 
for $f \le 1/24$.
Such loop excitations have
received far less attention in 3D systems\cite{feynman55} than their 2D
counterparts, possibly because they require more energy to excite and
therefore matter only very close to the mean field transition.
But
in high-T$_c$ materials, the short correlation lengths, high anisotropy, and
high T$_c$ broadens the vortex-loop-dominated
regime\cite{carneiro92,chatto94}, before amplitude fluctuations set in.

In a cubic sample with periodic boundary conditions, all
vortex lines naturally close
on themselves to form loops.  These loops are of two topological types:
those which can continuously shrink to a point (``trivial class'') and
those which cannot (``nontrivial'').  The latter are said to have a nonzero
``winding number,'' i.\ e., number of infinite lines in a given direction.
In the 2D periodic case,
the loops lies on the surface of a torus.
Here, there are two distinct subclasses of
non-trivial loops: one which winds around its circumference,
and another which runs transverse to it.
In the infinite 2D geometry, these correspond to
lines infinite in either the $\hat{x}$ or the $\hat{y}$ direction.
On the 3D hypertorus, there are infinite lines in any of {\em three}
directions.

These notions play critical role in the description of dissipation
via vortex motion, i.\ e.\ phase slips\cite{langer67_2}.
For current flowing in a given direction,
the dissipation may occur either through
the expansion of loops, or through motion of an infinite
line. In either case, the dissipation arises from
fluctuations in the winding number of
vortex lines perpendicular to the current(c.~f.~Equations (\ref{gameq}) and (\ref{sqcriteq})).
Note that when a finite field
is applied along $\hat{z}$ direction, the hypertorus already contains
many ``windings" along that direction even without an applied current.

In the absence of pinning, dissipation in the plane normal to $\hat{z}$ is
governed by fluctuations of the winding number in the $\hat{z}$ direction.
This dissipation should not depend directly on
whether or not the ``windings"  - that is, the vortex lines - form
a lattice, but may depend on the {\em density} and mobility of windings.  

Dissipation parallel to the field direction
(c-axis resistance) depends mainly on winding number fluctuations
transverse to $\hat{z}$.  Clearly, the average winding
in this direction vanishes,
unless the field-induced lines themselves, while
winding along $\hat{z}$ as required,
also wind along another direction like wires
around a solenoid. For this to occur, the lines
would have to break a chiral symmetry,
spontaneously generating a global surface current with a net magnetization
normal to $\hat{z}$ - an effect which should be prohibited energetically in the
ground state.

\begin{figure}[th]
\begin{center}\leavevmode\psfig{figure=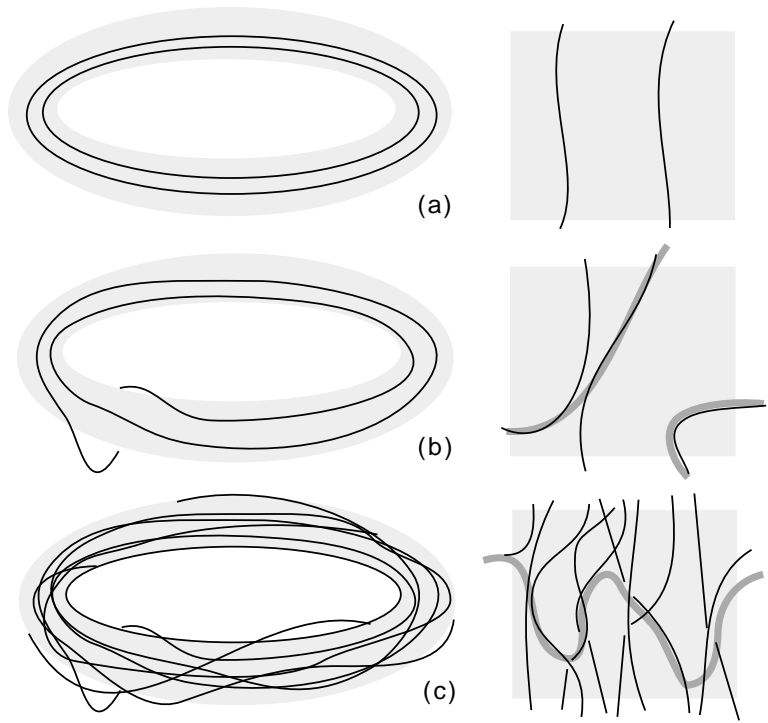,width=4in}\end{center}
\caption [lala] {Effect of switching connections among entangled flux lines
in a torus geometry [panel (a)] and in an infinite plane with open boundary
conditions. For the case of only two lines [(b)],
note that it is necessary to make a long
excursion spanning the whole plane to change the global winding number.
The latter may be easily induced in a dense environment through collective
occurrence of local reconnections [panel (c)].
}
\label{donutfig}
\end{figure}

It may occur, however, if there exist 
entangled field induced vortex lines which collide with each other to switch connections(a process
we may call ``cutting and reconnection'').
In Fig.\ \ref{donutfig}, we show two
field induced lines residing on the surface of
a torus (left panel) going through such a
cutting and reconnection [(panels (a)-(b)].
The right column of the figure shows an alternative
view of the same process in an infinite space with open boundary conditions.
Initially, both vortex lines
wind only along the $\hat{z}$ axis. After the cutting
and a special reconnection process in which one strand circles around the
torus before meeting its other end,
a net transverse winding number has been created.
This ``global'' process is, however, energetically expensive
because it involves a spatially extended excursion [panel (b)] and should
occur
very rarely, even in the melt.

If we introduce an ``entanglement length" $\ell_c$, defined as the average
distance  along $\hat{z}$ required for any two vortex lines to wind
around each other, we expect $\ell_c$ to be infinite for $T < T_m$, but to
become finite in the liquid phase.
Because of the finite line tension and repulsive interactions between vortex
line segments, such entanglement events along the flux lines are costly in
energy and hence rare, in the liquid near melting.  Deeper into the liquid phase,
as the repulsive interaction between vortex lines is overcome by entropic
forces of attraction, $\ell_c$ should become much
shorter, leading to a much denser entanglement pattern.
The now numerous local transverse
fluctuations, and local cutting and reconnection events
(i.\ e., collisions) generate
fluctuations in the ``global" transverse winding number and
cause $\gamma_{zz} = 0.$
On symmetry grounds, the average transverse vorticities
$< n_x > $ and $< n_y >$  have
to be zero at all temperatures. But possibly  $< |n_x|^2 +|n_y|^2 > $
acquires a finite value for $T>T_{\ell}$, suggesting that this quantity could
be used as another ``order parameter" for the hypothetical
phase transition at $T = T_{\ell}$ (with a nonzero value at
{\em higher} temperatures).

Alternatively, we may view the upper transition in the context of a bond percolation
transition.  The field induced lines provide  a kind of backbone network.  With
increasing $T$, vortex lines undergo more and more transverse collisions.
At $T = T_{\ell}$, these collisions induce the entire ensemble of
field-induced vortex lines to form an infinite
connected  $D-1$ dimensional structure transverse to the applied field, 
causing large fluctuations
in the transverse winding number [thick gray line in panel (c)], thereby wiping out any  
superconducting path connecting the top and the bottom layers normal to the field.

Let the mean-square transverse displacement of field-induced vortices per layer
be denoted $\ell_T^2\equiv < |{\bf r}_i(z) - {\bf r}_i(z-1)|^2 >.$   Then
$\ell_c / d$ is defined as the number of layers
along $\hat{z}$ over which a line wanders transversely by the
average intervortex distance.  We write this condition as
$\ell_T^2 \cdot [\ell_c/d]^{2\zeta} = a_B^2$,  where we introduce
an unspecified ``wandering" exponent $\zeta$.  In the
limit of dilute (independent) lines, we expect $\zeta \approx 1/2$,
corresponding
to a random walk of each vortex line segment.
Long-range intervortex repulsion is known to renormalize
the unit step size $\ell_T$ from $c(T) \cdot (T/J_z)^{0.5}$
down to a smaller value with a similar form with an unspecified T dependence encoded in $c(T) < 1
$\cite{monica94,ryu_thesis}.  It is not clear how $\zeta$ is affected by
intervortex repulsion, but possibly
the interactions with other fluctuating lines are
equivalent to the line of interest being in a random environment.
For a flexible line in a 3D random environment, $\zeta \sim 0.6$
\cite{kardar87,balents93}.
For $D \ge 2$, in the actual system of interacting fluctuating lines, no exact result is available
for $\zeta$, although several numerical results
and conjectures give $\zeta \sim 0.2-0.6$\cite{krug}.

We can use these crude estimates to make a guess at the field dependence
of $T_{\ell}$, interpreted as a bond percolation transition.
Along a given field-induced vortex line,
the probability per unit length
that a transverse connection is made to a
neighbouring vortex line at any position along the $\hat{z}$-axis
is $p = d/\ell_c = (\ell_T / a_B )^{1/\zeta}.$ Since $\ell_T \propto (T/J_z)^0.5$ and 
$a_B \propto B^{-0.5},$ 
the percolation threshold is reached roughly
when $ c(T)^2 TB/J_z > [p_c]^{2\zeta}$, 
where $p_c$ is an appropriate percolation
threshold.  This condition defines
a {\em lower bounds} for a possible transition at $B_{\ell}(T)$ which approximately follows
\begin{equation}
\label{dceq}
B_{\ell} \propto { [p_c]^{2\zeta} \over c(T)^2}{J_z(T) \over T}.
\end{equation}
For a dense lattice or large anisotropy (i.\ e., small $J_z$),
this condition is probably satisfied
immediately upon melting, as at $f=1/6.$
For dilute systems, however,
the second transition is not automatically triggered by melting
and may occur only deep into the liquid phase, at a temperature where
the entanglement barrier is sufficiently weak to allow an infinite vortex
tangle to form.  Whether this percolation transition is a true phase transition
or only a sharp crossover remains to be determined.

This same picture hints at how correlated pins
such as columnar damage
tracks\cite{civale91} may increase $T_{\ell}$.
Such columnar disorder
will encourage the vortex lines to stay straight along the defect
track, reducing the effective unit step $\ell_T$
by a factor $c_p \ll c.$   As a result, the
wandering exponent $\zeta_p$ may also change from its thermal value $\zeta$.
Consequently, $T_\ell$ will be enhanced by an overall factor of
$({c \over c_p})^2({1 \over p_c})^{2(\zeta-\zeta_p)}$.

In summary, the upper transition is characterized by
the following set of equivalent
criteria:\cite{li93_2,jagla96} Disappearance of finite transverse diamagnetism;
disappearance of phase rigidity along the field direction;
appearance of an infinite {\em transverse} vortex cluster;
large fluctuations in the global transverse winding number or the net
vorticity ${\cal M}$; onset of finite c-axis phase-slip resistance in the
limit of vanishing bias current in the $c$ direction, which is equivalent to saying no 
superconducting path exists over macroscopic distances.

\subsection{Dissipation for $f=1/24$}

\label{f24rvstsection}

While Bitter decoration serves as a detailed probe of spatial vortex
configurations\cite{kim96,murray90,yao94}, it yields ambiguous
information about freezing, and is restricted to very low flux densities.
Cubitt {\em et al} obtained evidence of a
melting transition in \bscco
from low angle neutron diffraction\cite{cubitt93}.
Similar results were obtained by a $\mu$SR technique, which
probes the local magnetic field distribution\cite{lee93}.
NMR\cite{recchia95} and atomic beam\cite{harald} techniques have also
been used to study both the static properties and the melting
of the vortex lattice.
More recently, Schilling {\em et al} employed a differential thermometry to
search for the latent heat of melting in \ybco and to obtain a melting
curve\cite{schilling96,schilling97}.

Far more information has been accumulated from
transport measurements, but this is much less easily interpreted
in terms of vortex lattice melting.
The interpretation is complicated by disorder, as well as by
the fact that the measurements are nonequilibrium and usually involve
nonuniform current distributions.
Safar {\em et al}\cite{safar92} measured a sharp jump in resistivity in the
mixed state of \ybco.  The resistivity also showed a hysteretic behavior
upon thermal cycling, indicating a first order transition.
The transition line thus obtained seems to coincide with
``melting curves" obtained by torque measurements\cite{farrell91},
and more recently, by differential specific heat
measurements\cite{schilling96}.
Kwok {\em et al} have carefully demonstrated the effect of twin boundary
pinning on the melting transition in a series of
transport measurements which track the so-called
``peak effect" associated with vortex lattice softening \cite{kwok94}.
They find that the peak effect sets in at a few degrees below the
melting curve determined from a sharp kink in resistivity.
This sharp resistivity kink, as observed by
both Safar {\em et al}\cite{safar92} and
Kwok {\em et al}\cite{kwok92}, tends to become less pronounced both at 
very high$(B > 10 \, {\rm tesla})$ or low flux
densities$(B < 1 \, {\rm tesla})$\cite{sharpjump}.

An ideal, but impractical, transport measurement to determine the melting curve
would consist of applying an infinitesimal
current to induce a net Lorentz force on the lattice, which is held in
place by a balancing pinning force.
As soon as the lattice melts, individual lines would begin to drift,
inducing ``flux flow" resistance.
Most real materials, however, are complicated by disorder, and
even the static properties of the lattice with
disorder are incompletely understood\cite{bouchaud92_3,giam95}.  In the
presence of disorder, varying the field density produces changes in
both the effective pinning strength and the effective flux lattice anisotropy.
Depending on relative strengths of all these competing effects, many
complications may arise in probing thermodynamic properties using transport
experiments\cite{danna95,koshelev94,hellerqvist96,jensen96,ryu97}.

A number of recent transport measurements have, nonetheless, produced rich
information about flux lattice melting.
Pastoriza {\em et al} have used
a non-uniform distribution of pinning strength to
probe the shear modulus directly\cite{pastoriza95}, giving direct
information about the lattice stability.
Zeldov {\em et al}\cite{zeldov95} have used local Hall probes in \bscco
to monitor the local field density.  They found
a very sharp jump, again interpreted as a signature of a first order
melting transition.  More recently\cite{fuchs96},
Fuchs {\em et al} performed simultaneous measurements of the resistance
and local magnetization, confirming that that the jump in local magnetic
density
coincides with a sharp increase in  resistance (but not necessarily a
discontinuous {\em jump}).
The heat of melting {\em per vortex per layer} inferred from this local
magnetization jump, however, shows rather peculiar features:
it vanishes continuously as the field is increased,
while steeply increasing as the field is lowered toward zero.

These results raise
several outstanding questions: How can the seemingly first order
transition line terminate apparently at a point in the H-T plane?
Does the melting line monotonically approach
$T_c(H=0)$ or follow a reentrant melting curve?
Another important issue is the longitudinal phase coherence probed
by c-axis resistivity vs. T measurements\cite{briceno91,gray93,monica94},
which show a striking series of broad peaks in \bscco single crystals.
The nonlocal conductivity associated with this phase coherence can be
probed in the so-called flux transformer geometry. The experimental data of 
Keener {\em et al}\cite{keener96} showed that phase coherence over 
a finite correlation length along ${\bf B}$ persists above the melting 
transition of the vortex lattice in some region of the H-T phase diagram.

A simple and natural model for probing the dynamics of the mixed state
is a network of resistively shunted Josephson junctions with Langevin
noise.  In this section, we present some results of simulations using this
model, and to connect these to the analogous static XY results.
Our calculations are carried out as follows. At any given temperature,
the final snapshots from the MC simulations are used as the initial
dynamical phase configurations.
We use an integration time step $\triangle t = 0.1t_0$.
After the current is switched on, $1000-5000 \triangle t$ is allowed
for the system to reach a steady state, following which
the voltage is averaged over the next $6000- 12000$ steps of $\triangle t$.

\begin{figure}[th]
\begin{center}\leavevmode\psfig{figure=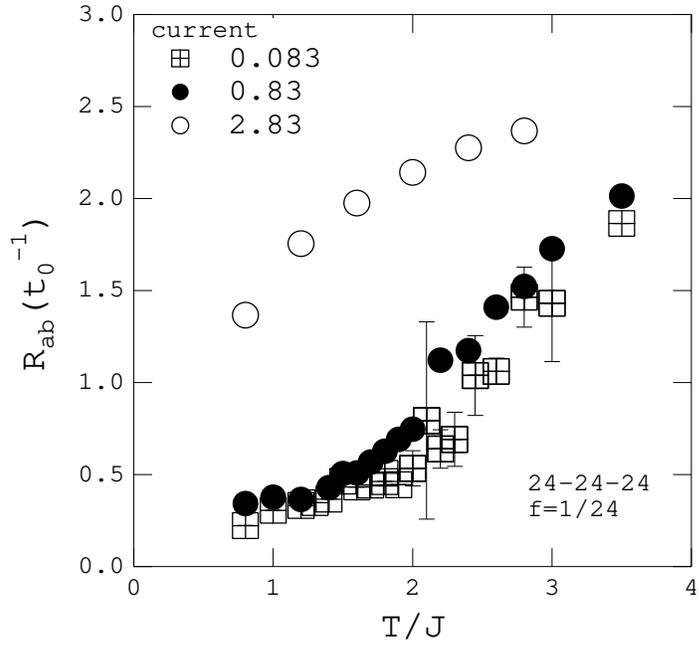,width=4in}\end{center}
\caption [lala] {Calculated bulk in-plane resistance vs.
$T$ for $f=1/24.$
Currents of $0.083 - 2.83 I_c$ per grain
were injected uniformly into an $yz$ plane.
}
\label{rvstfig}
\end{figure}

Fig.\ \ref{rvstfig} shows the ``bulk in-plane resistance'' at $f = 1/24$.
The measurement geometry is as shown in Fig.\ \ref{geomfig} (a); thus,
these calculations probe the shear rigidity of the lattice in contrast to the
usual transport experiment in which random pins play an essential and
complicating role.
Through the rest of this paper, we will call these calculated quantities 
$R_{ab}$ and $R_c$, even though they differ
from the resistivity measured in most transport experiments.
At the highest bias current of 2.83$I_c$ per grain(equivalent to
$1.4 I_c$ per bond), we have a smooth curve without any noticeable
changes either at $T_m$ or at $T_\ell.$
For lower values of driving current, sharper features emerge. There is a
slope discontinuity near $T_m = 1.5 J$ for both $I/I_c = 0.83$ and $0.083$, but
the most dramatic change occurs near $T_\ell = 2.0 J$.
The entangled line liquid for $T_m < T < T_\ell$ seems to have a sizable
viscosity.  This viscosity impedes the motion of the flux lines in the
liquid, the two halves of which are driven past each other by opposing
Lorentz forces.  As a result, the lines move slowly, and dissipation (defined
as a ``resistance'' R$_{ab}$) is small.
The steady increase of $R_{ab}$ with temperature in this region is due to
screening by vortex loops which gradually lowers the viscosity.
For  $T>T_{\ell},$  this viscosity vanishes, leading to
a steep increase in $R_{ab}$   This increase at
$T_\ell$ is enhanced by  additional (and
probably dominant) dissipation produced
as the system goes through an XY-like transition or crossover.
The large viscosity for $T_m < T < T_\ell$ is also
consistent with the slow ($\ln t$) equilibration seen
in the Monte Carlo measurement of $\gamma_{zz}$
for $T_{m} < T < T_{\ell}$\cite{logt}.
We believe that the change in $R_{ab}$
near $T_\ell$ shares the same mechanism as that seen
near $T_{XY}(f=0)$ shown in Fig.\ \ref{f0rvstfig}.

\begin{figure}[th]
\begin{center}\leavevmode\psfig{figure=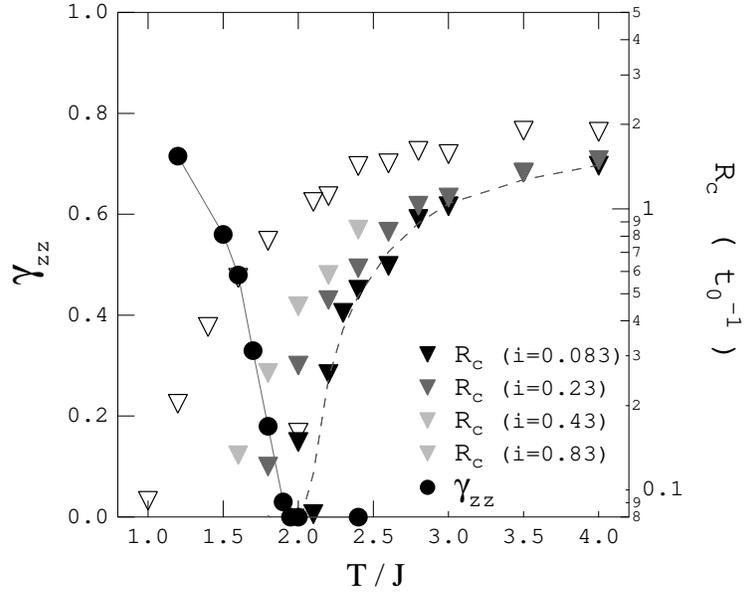,width=4in}\end{center}
\caption [lala] {Calculated $c-$axis resistance (arbitrary units) vs. $T$
for $f=1/24.$  A current of 0.083I$_c$ per grain
is injected uniformly into a plane.  The inset shows
the same results, but with temperature rescaled to model
a hypothetical {\em isotropic}
high-$T_c$ superconductor as in the previous Figure.
}
\label{caxisfig}
\end{figure}

Fig.\ \ref{caxisfig} shows the ``$c$ axis resistivity" $R_c$ at f = 1/24,
as calculated using the geometry of Fig.\ \ref{geomfig} (b).
For comparison, we also show the calculated
$\gamma_{zz}$.   As the driving current is reduced, $R_c$ seems
to approach a curve which vanishes asymptotically as $T\rightarrow T_\ell^+$,
coinciding with the vanishing $\gamma_{zz}.$
Our numerical results thus suggest that the dramatic increase in $R_c$
results from the $T_\ell$ transition rather than melting.
The increase in $R_c$ is thus correlated with
massive vortex line cutting, as put forward 
earlier\cite{ryu_thesis,monica94}, and with an increase in the
density of transverse vortex segments $<|n_{xy}|>$,
the frequency of vortex line crossings, and
fluctuations in the transverse net vorticity, $\delta {\cal M}_{z}^2$.
This distinction between $T_{e\\}$ and $T_m$ may be most important for \bscco
at low fields, and in disordered dense systems\cite{jagla96},
where the temperatures may be most separated.

We can draw some conclusions relevant to experiment from the current dependence of both $R_{ab}$
and $R_c$ observed in our simulations.  
First, the ``melting line,''
{\em as detected via a voltage criterion at constant current}
in a bulk resistance measurement, should
be sensitive to the driving current, even at a very low bias.  
Existence of pinning force is essential in getting distinct transport behaviors  
for the lattice and the liquid. 
On the other hand, the transition at
$T_\ell$, whether monitored by the vanishing of $R_c$  in the limit of
small current or by a jump in $R_{ab}$ between two {\em finite} values,
should be relatively insensitive to applied current density, since the main mechanism
of dissipation(presence of transverse vorticity) in this case is switched off below
$T_\ell$ and sets in above $T_\ell$ irrespective of whether we have pins or not. 
Indeed, just such an
observation has been made by Keener {\it et al}\cite{keener97} in describing their
curves for
$T_m(H)$ (melting) and
$T_D(H)$ (``decoupling transition''), as obtained by flux-transformer measurements on
\bscco single crystals. It is plausible that their  $T_D(H)$ at very low fields ($B <
100$ Gauss) corresponds to $T_\ell$ in our model. True melting line is presumably the
limiting value of the current-dependent $T_m(H, J)$ as $J\rightarrow 0.$ It is not 
experimentally verified whether such a limiting value coincides with $T_D$ or not.

Let us briefly comment on the experimental possibility of distinguishing
between $T_m$ and $T_{\ell}$
If we imagine a hypothetical {\em isotropic} high temperature superconductor,
the coupling constant $J$ in our model calculation is related to
the parameters of the superconductor via
$J \sim {d\phi_0^2 \over 16 \pi^3 \lambda^2(0)}
\times ( 1 - [T / T_{c0}]^4)$ (assuming the two-fluid model).
Taking $T_{c0} = 92 K$,
$d = 10 \AA$, and $\lambda(0) = 1000 \AA$,
we find $T_m \sim 89.7 K$ and $T_\ell \sim 90.3 K$ [eq. (11)].
Thus the two transitions are  remarkably close even for the isotropic
case.   In real materials such as
\ybco and \bscco, the separation between the two will be further reduced
by an anisotropy factor, although pinning disorder may tend to separate them.
Therefore, in many cases, it will be nearly impossible to separate the
two transitions experimentally.

\section{Local magnetization jump and heat of melting}
\label{zeldovsec} 

A striking result of the \bscco micro-Hall probe measurements
is the sharp jump in local magnetization (vortex density)
across the phase transition\cite{zeldov95}.
At ``high'' fields ($\sim 200 G$), the jump occurs at nearly constant $T$,
and even for lower fields, still within $\delta T \sim 3 mK.$
The heat of melting per vortex per layer,
$T_m \triangle S = - {T_m \triangle B \over 4 \pi} {dH_m \over dT}$,
as obtained from the Clausius-Clapeyron relation,
increases monotonically from $0$ at $B \sim 400 G$
to about $0.6 k_B$ at $B \sim 55 G$, beyond which the slope
${dH_m \over dT}$increases very sharply (cf.\ Fig.\ 6 of \cite{zeldov95}).

Similar jumps also seem to occur in \ybco, as reported in
recent calorimetric\cite{schilling96}
and magnetization measurements\cite{liang96,welp96}.
The estimated latent heat of melting
yields $\triangle S$ (per vortex
per layer) $\sim 0.4 k_B$ for $1 < B < 8$(tesla).
The data (cf.\ Fig.\ 1 of Ref.\ \cite{schilling96})
shows that the jump $\Delta M$ at $T=85 K$ ($B\sim 3.7$ tesla) 
for \ybco is spread over a
field range $\delta B \sim 0.1T$, or, for a given field,
over a temperature range $\delta T \sim 0.1 K$.
This jump decreases rather abruptly for flux densities
$B \leq 1 T$.  The estimated entropy of melting
($\sim 0.4 k_B$ per vortex pancake) is quite close to that
numerically obtained by Hetzel{\em et al}\cite{hetzel92} ($\sim 0.3k_B$ per
pancake), and also to the values obtained in model calculations
based on the lowest Landau level
and London approximations\cite{errata}.

As the field decreases, the jump in $M$ occurs over a
broader temperature range (cf.\ Fig.\ 3 of \cite{zeldov95}).
The resistance jumps measured by Kwok {\em et al}\cite{kwok92},
attributed to the melting transition,
also become broader with decreasing field.
By contrast, the height of the resistivity kink seems quite
uniform over a wide range of fields.
These last two observations are consistent, however,
if we interpret the jump in resistance as a signal that
$\gamma_{zz} \rightarrow 0$.
In view of all these facts, it is plausible that, at least at
relatively high fields which corresponds to $f J_{xy} / J_z > 1/18$,
the experimental jumps observed in local
magnetization and resistance\cite{zeldov95,schilling96,kwok92}
shows the combined effects of two distinct processes,
occurring within $\triangle T < 10 mK.$  As a corollary,
the very low field measurements($f J_{xy} / J_z < 1/18$) 
actually may not track the
melting transition itself, but various manifestations of the predominant XY
fluctuations (vortex loops)
which are most conspicuous near $T_\ell.$.

Observation of the {\em reentrant} melting curve has 
not been reported in any high-T$_c$ materials.
In NbSe$_2$, the melting line detected by the peak effect was reported to
be non-monotonic in field\cite{sabu}, consistent
with the reentrant melting curve  proposed for
the more anisotropic high-T$_c$ superconductors\cite{nelson88}.
But if one interprets the magnetization jump in \bscco as evidence for melting,
then the melting curve for \bscco apparently
approaches $T_{c0}$ {\em monotonically} at field as low as $\sim 1 G$.
This behavior is surprising since, at these fields, the
vortex separation far exceeds the magnetic screening length.
Furthermore, in \ybco, the peak effect at $0.35 -1.5$ T
is observed to lie below the resistivity kinks\cite{kwok94}
(about $0.8 K$ below at $0.5$ T).  These measurements suggest
that, at least
for low flux densities, transport measurements may actually
not be probing flux lattice melting.

We propose that low-field melting
is indeed reentrant.  Most low-field experiments which probe
magnetization\cite{zeldov95,fuchs96},
thermal properties\cite{schilling96}, and transport coefficients
\cite{safar92,fuchs96,keener96,kwok92} actually
track $T_{\ell}$, which is progressively more
separated from $T_m$ and approaches $T_{XY}(0)$ as the field is reduced.
We have already shown that dramatic changes in $R_{ab}$ and $R_c$
occur at $T_{\ell}.$   Moreover, the broad peaks in $C_V$ are centered at
$T_\ell$ and they, too, approach $T_{XY}(0)$ as $B$ decreases.
Our estimated upper bound for the total
entropy release in the temperature range $T_m <  T < T_\ell$,
as estimated from the $T-$dependence of the internal energy, 
qualitatively resembles that of
Zeldov {\it et al} (Fig.\ 6 of \cite{zeldov95}) in that it steeply increases
as $f$ decreases.
At higher fields, the observed (and also
calculated) $\triangle S \sim 0.3-0.5 k_B$
is consistent with the destruction of phase coherence
at a {\em single} transition.
By contrast, at low fields, phase rigidity is lost in a two-step
process.  Most of the entropy release
($\triangle S \geq 0.5 k_B$ per vortex per layer) occurs near $T \sim 
T_\ell > T_m$,
whether or not this is a true phase transition.
Note that a hysteresis in the resistivity may be observed near $T_{\ell}$ 
due to finite vortex-cutting barriers below
$T_\ell$.  This is not necessarily an evidence for a first-order melting
transition at very low fields.

The melting line in the dilute limit may be quite difficult
to detect experimentally.
Conceivably it may be tracked by the peak effect, by high-resolution
IV measurements, or by direct measurement
of the shear modulus \cite{pastoriza95}.
Of course, direct observation of a vanishing neutron diffraction pattern
as in \cite{cubitt93} would be ideal, but this technique
is of limited applicability in this density range.

To shed further light on this problem,
we have carried out calculations {\em with mixed boundary conditions}.
That is, we allow local density fluctuations in the net z-component of
vorticity by using free boundary conditions in $x-$ and
$y-$ directions, while retaining periodic boundary conditions
along the $z-$axis.  Of course, surface effect are now stronger, possibly
reducing the melting temperature.  Another point is that
our uniform-frustration model assumes
that $\lambda = \infty$.  Therefore, we should proceed with some caution in
relating our numerical results to experimental data.

To study the system with these mixed boundary conditions, we again
did a simulated annealing run for a single layer of the triangular grid
to find the lowest energy configuration.
By stacking the resulting state layer
by layer, we form the ground state lattice, which, because of
incommensurability and the free boundaries, now
consists of an imperfect triangular lattice with some defects.
This lattice melts at $T/J \le 1.4$ for 
the nominal density of $f=1/24$ on a $26\times 26\times 12$ grid.   This is
slightly below the value $T_m/J \sim 1.5$
found for the fixed density system of 24 layers with periodic boundary
conditions.  For a nominal density of $f=1/6$,
melting occurs near $T/J \le 1.15$ with these mixed boundary conditions.

One might think of defining the
``magnetization'' $M_z$ as the average net vortex
density $n\equiv \int n_z({\bf r}) d{\bf r} / A$,
where $n_z({\bf r})$ is the local vortex density and $A$ is the total area.
However, $M_z$ defined in this way suffers from
spurious boundary effects, arising from
the depletion of vortices near the boundaries in the lattice phase\cite{ebner}.
Upon freezing, the lattice develops a {\em rigid} free surface of irregular
shape, expelling some of the vortices from the rectangular bounding box.
The resulting change in density, $\delta n / n$, is an artifact of the
open boundary conditions, and we find that it vanishes for
large samples as $1 / \sqrt{A}$, confirming that it originates from a surface effect.

\begin{figure}[th]
\begin{center}\leavevmode\psfig{figure=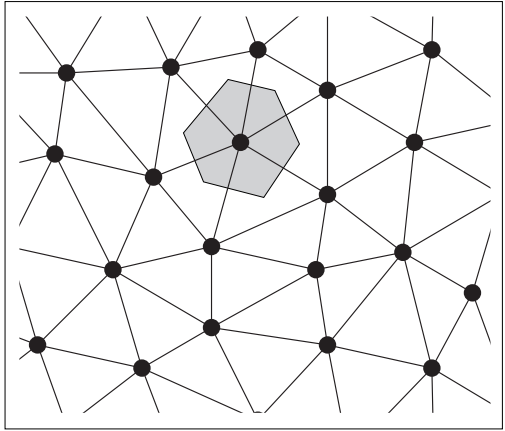,width=3in}\end{center}
\caption [lala] {Example of the local Voronoi cell occupied by a vortex.
}
\label{vorfig}
\end{figure}

\begin{figure}[th]
\begin{center}\leavevmode\psfig{figure=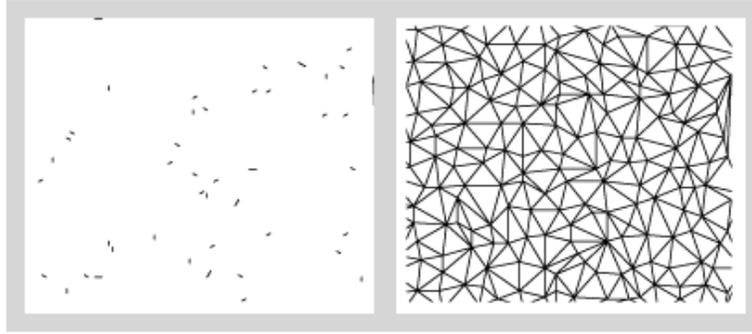,width=4in}\end{center}
\caption [lala] {Example of the vortex configuration in a xy-plane at
$z=6$ in a 48-48-12 system with $f=1/24$ at $T_\ell.$
The black dots are field induced vortices, gray dots connected by
lines are bound dipoles identified. The right panel shows the
Delaunay triangulation applied to the field induced vortices only.
}
\label{dipolefig}
\end{figure}

Instead, we define $M_z$ by a criterion involving
the {\em local Voronoi cell area} ${\cal A}_i$, i.\ e.,
the area of the generalized Wigner-Seitz cell for
vortex $i$ (the shaded area shown in Fig.\ \ref{vorfig}).  Before applying
the procedure, we first eliminate the thermally induced vortex loops, which
are present in addition to  the field-induced vortices for
$T \ge 0.5 \times T_\ell$.  To do this, we pair each antivortex with the
nearest vortex in each plane, identifying the resulting pairs as bound dipoles
to be excluded from the count (cf.\ left panel of Fig.\ \ref{dipolefig}).
Since most such dipole pairs have linear dimensions much smaller than
$1/\sqrt{<n>}$, this criterion is justified.
We then perform a Delaunay triangulation on the field induced vortices
to determine topological neighbors for each vortex.  From the bond
configuration thus determined, we obtain its dual, which is the desired Voronoi
diagram.  A local vortex density at a point ${\bf R}$
may then be defined as
\begin{equation}
\label{localdensityeq}
n({\bf R}) = \sum_i \delta( {\bf R} \in {\cal A}_i )
 / {\cal A}_i
\end{equation}
where $\delta ({\bf R}  \in {\cal A}_i) = 1$ if the
point ${\bf R}$ lies in the Voronoi cell associated with vortex $i$, and zero
otherwise.   Next, the local magnetization $M_z$, which we interpret as
the {\em bulk average density} $<n>$ is calculated from
\begin{equation}
<n> = \sum_{{\bf r}_i \in {\cal C}}  { 1 \over {\cal A}_i },
\end{equation}
i.\ e., as the average of the inverse Voronoi area for vortices lying within a
measurement area ${\cal C}$ suitably distant from the sample boundary.

\begin{figure}[th]
\begin{center}\leavevmode\psfig{figure=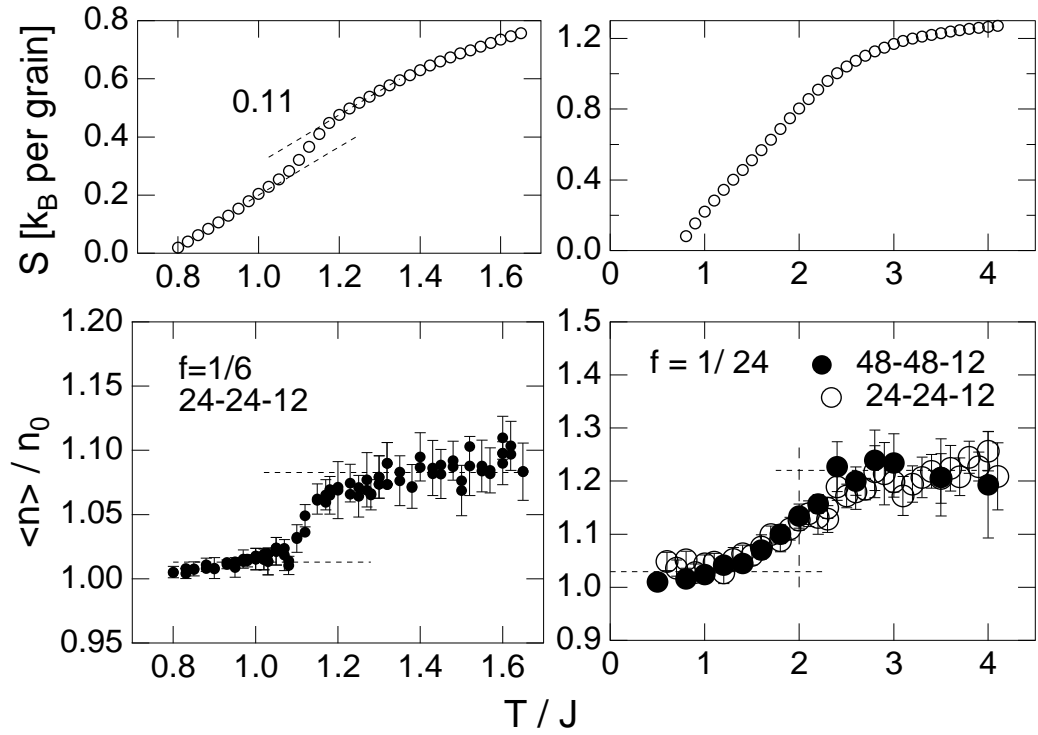,width=6in}\end{center}
\caption[lalal]{Entropy $S$ (upper panel)
and normalized vortex density (lower panel) for $f=1/24$ and $1/6$ in
$24 \times 24 \times 12$ and $48 \times 48 \times 12$ systems with open
boundary conditions. The error bars denote the rms deviations from layer
to layer.}
\label{jumpsfig}
\end{figure}

In Fig.\ \ref{jumpsfig}, we show the relative average vortex density (filled
circles) along $\hat{z}$, $<n_z> / n_0$, normalized to the
nominal density per layer at $f=1/6$ and $f=1/24.$
For $f = 1/6$, $<n_z>$ shows a sharp jump at $T_m$ to a value about
7 \% larger than $n_0$.
For $f=1/24,$ the there is a similar change of about 15 \% which is less
sharp than at $f = 1/6$ and is centered at $T_\ell.$
From these two data points, we observe that
$[<n>(T_\ell) - n_0] / n_0 \sim
f^{-1/2}.$
Comparing the result for two different sample areas ${\rm L}_x{\rm L}_y$ for $f=1/24$,
we have verified that the observed change is indeed a {\em bulk} phenomenon,
independent of any surface influence.
Note that we have about the same number of flux lines($\sim {\cal
O}(200)$) for both $f=1/6$ and $f = 1/24$ for our chosen sample sizes of
$24 \times 24 \times 12$ and $48 \times 48 \times 12$.
While the jump occurs at $T_m$ for $f=1/6$, we do not observe a similar
feature near melting ($T_m = 1.35 J$) at $f=1/24.$ Therefore, the cause of
the jump in the  local vortex density should be sought in the nature of
transition at $T_\ell$, rather than in the mechanism for
flux lattice melting.
Note that these jumps resemble those in
Fig.\ 5 of \cite{zeldov95} in the ``anomalous low-field regime'' $(1 < B < 55G)$
in the following sense: the
fractional change in vortex density decreases, and the jump becomes sharper, 
as the field increases.  Probably, the line density 
$\langle n_z \rangle$ increases
with increasing T for
$T_m < T \le T_\ell$ because the repulsive intervortex
interaction is screened by polarizable
vortex loops.  
The 2D analog of this effect is the screening of the repulsion between
field-induced vortices by thermally excited vortex-antivortex 
pairs\cite{doniach79}.

Does the jump in flux density occur exactly at the melting transition, or
is it more closely connected to the other ``transition'' at
$T_\ell$, i.\ e., to a transition between two liquids with different
compressibilities?  This question may actually be rather academic,
since $T_m$ and $T_{\ell}$ may practically merge in real, anisotropic
materials at high fields.  Eq.\ (\ref{dceq}) gives a rough criterion for $T_{\ell}$ in
isotropic systems: $1/\ell_c = (c^2 BT/J_z)^{1/\zeta} > p_c$.
For anisotropic systems such as \bscco and \ybco, a given value of $f$
corresponds to a field which is reduced, relative to the isotropic system, by
a factor of $J_z / J_{xy}.$  Therefore, the merging of $T_{\ell}$ and $T_m$,
which in isotropic systems occurs around $f \sim 1/18$,
should in anisotropic materials occur around f =  (1/18)$J_z/J_{xy}$,
This anisotropy factor $J_z / J_{xy}$ could be as small as ${\cal
O}(0.0001)$ in \bscco.

\section{Discussion}
\subsection{Analogy to XY Transitions of Slabs of Finite Thickness}

In the previous sections, we made following observations from numerical simulations:
(i) $\gamma_{zz}$ vanishes at $T_\ell\ne T_m$ for isotropic system with $f >
1/18$ and $T_\ell(B)$ appears to terminate at
$T_{XY}$ for $B \rightarrow 0$;  (ii) As $B$ changes, it
tracks the broad peak in $C_V$ which behaves 
\begin{equation}
\label{cpeakeq}
C_V^{max}(B) \sim - \ln (L_B),
\end{equation}
and 
\begin{equation}
\label{deltatceq}
|T_{\ell} - T_{XY}| \sim L_B^{-2/3},
\end{equation}
where $L_B \equiv f^{-1/2} \sim B^{-1/2}$; (iii),
$T_{\ell}$, and not $T_m$, appears to coincide with the
principal change in local vortex density which follows
\begin{equation}
\label{deltameq}
\triangle M / B \sim  B^{-1/2}.
\end{equation}
There is a corresponding change in bulk resistances $R_{ab}$ and $R_c$ over the same 
temperature range;
(iv) $T_\ell$ and $T_m$ seem to merge at a sufficiently high field.

\begin{figure}[th]
\begin{center}\leavevmode\psfig{figure=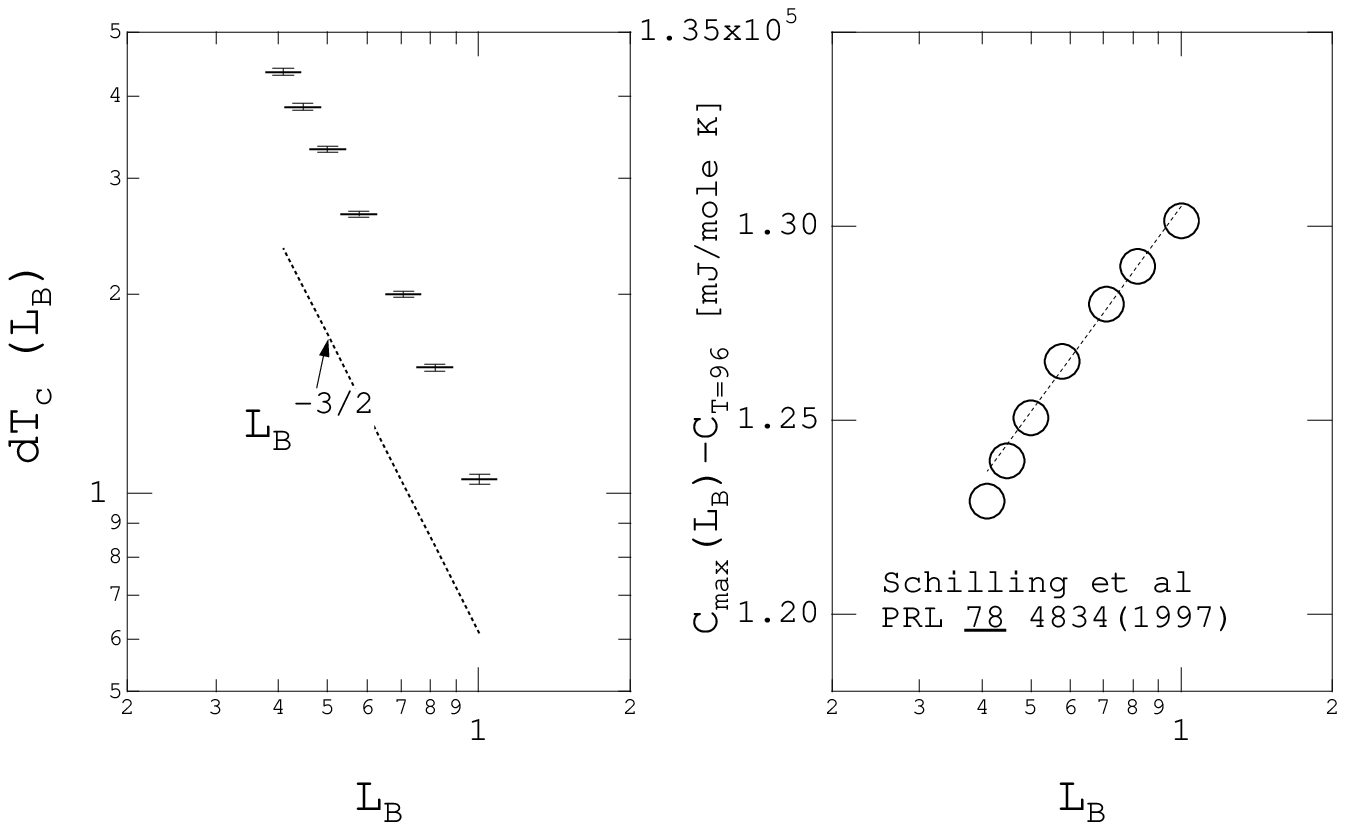,width=6in}\end{center}
\caption[lalal]{Dependence of ``broad" peak maximum in specific heat
and their shift in temperature  on the magnetic length $L_B$ taken from
Schilling et al\cite{schilling97}.}
\label{schillingfig}
\end{figure}

We now summarize some recent experimental observations which appear to be
consistent with these numerical results.
Schilling et al\cite{schilling97} have reported high resolution
calorimetric evidence for a first order transition in \ybco, which they
interpret as a melting transition.  They observe a very sharp delta-function
like peak lying on the left shoulder of the broad peak in
specific heat in the range of $0.75 - 9 $ tesla, which roughly corresponds to $1/81 <
f < 1/6$ in our isotropic sample. The delta-function
appears to vanish for densities lower than about 0.5 tesla. 
This remarkable experiment thus establishes the
existence of a first order melting transition line, which empirically follows
$|T_m - T_c(0)| \sim L_B^{-1.61}$ over $0.75 - 9 $ tesla.
These data are consistent with our numerical results on the following points:
(i) a first order melting transition exists, and becomes weaker
as the flux density is lowered; (ii) the melting transition is located on
the left shoulder of a broader peak in $C_V$; and
(iii) the height of this broad peak and its position generally follow the
behavior described in equations \ref{cpeakeq}, \ref{deltatceq} and \ref{deltameq}.
As further evidence of the correspondence, we show in
Fig.\ \ref{schillingfig}, the
data extracted from Fig.\ 1 of \cite{schilling97} for fields as high as
6 T.  At higher fields ($>7$ T), the points deviate from the observed power
law behavior shown in the Figure.

Another experimental data which agrees well with our numerical results is that
of Roulin et al\cite{roulin96}.  These workers have reported
that both the melting curve $T_m(H)$ and a point they label the
``superconducting-normal'' (SN) transition as monitored by
tracking the maximum in $C_V$ as a function of $H$
both follow the equation
$|T - T_c(0)| \sim L_B^{-3/2}$, consistent with our numerical results and
the scaling analysis discussed above (to within logarithmic corrections).
Welp {\it et al}\cite{welp96} have reported
a detailed study of $\Delta M$ as a function of $T$ and $H$.
The data presented in
Fig.\ 2 of their paper shows that $\triangle M / B \sim B^{-1/2}$ for $1.8
\le B \le 5.6$ tesla, once again in agreement with both our numerical results
and the scaling data  From this data, we conclude that our
numerical observations, based on a frustrated 3D XY model, are generally
consistent with recent experimental observations.

Most interpretations of these experimental results focused on only one true phase
transition in the low field regime, namely a first-order liquid-solid
transition. This viewpoint is consistent with our numerical results only if we 
assume that 
$T_{\ell}-T_m$ (where $T_{\ell}$ is defined as the temperature where
$\gamma_{zz}$ vanishes) will go to zero in the thermodynamic limit.
In the following, we will briefly review a multicritical scaling approach which
assumes a single melting transition which happens to  be in the  vicinity of the
zero field XY critical point. Our data are not sufficient to
determine without ambiguity whether or not this assumption is correct.
Therefore, we follow it by giving an
alternative discussion based on the hypothesis that there are actually two separate
phase transition lines:
$T_m(H)$ for flux lattice melting and $T_\ell(H)$ for complete destruction of
any superconducting path (phase coherence) in all directions.

Friesen and Muzikar\cite{moloni96} describe the 
SN transition at a finite ${\bf B}$ in the vicinity of the $f=0$ XY critical point.
Their scaling hypothesis takes the form
\begin{equation}
\label{scaleq}
f_s(B,T) \sim |t|^{2-\alpha} \phi_\pm (B |t|^{-2\nu})
\end{equation}
for the singular portion of the free energy density
in the XY critical region. 
Here $\alpha$ and $\nu$ are the standard critical exponents
describing the specific heat and correlation length of the $f=0$ critical point,
$t = T-T_{XY}(0)$, and
$\phi_{\pm}$ are appropriate scaling functions.  
$B$ is put in by hand based on the assumption that 
it is the only relevant length scale. It is plausible, but does not have rigorous 
justification.  Since $\alpha \sim 0$ and
$\nu \sim 2/3$ for the $d = 3$ $XY$ model, this expression can be rewritten
as
\begin{equation}
\label{scaleq1}
f_s(B,T) \sim |t|^2\ln |t|\phi_{\pm}(B|t|^{-4/3}).
\end{equation}

The singular part of $C_V \sim -\partial^2f_s/\partial t^2$ can now be
shown to satisfy the relation (for $T < T_{XY})$
\begin{equation}
\label{cveq}
C_V(B, T) \sim {\cal C}(x)\ln t
\end{equation}
where $x = B|t|^{-4/3}$ is the appropriate scaling variable and
${\cal C}(x)$ is another scaling function.  From this
we find (i) that the quantity $C_V/\ln |t|$ has a maximum at some fixed
value of $x$; and (ii) at that fixed value of x, the maximum value
$C_v^{max} \sim \ln |t|$.  Both (i) and (ii) are in agreement with our
numerical data.  Similarly, the magnetization is given by
$M(B,T) \sim (\partial f/\partial B)_T$.  It is readily shown to satisfy
\begin{equation}
\label{mageq}
M \sim {\cal M}(x)(B^{1/2}\ln |x| - \ln B),
\end{equation}
where ${\cal M}$ is another scaling function.  A reasonable interpretation
of the ``jump'' $\Delta M$
in magnetization is the difference in M between two fixed
values $x_1$ and $x_2$ of the scaling variable.  Then, if the term involving
$\ln B$ can be neglected, we have $\Delta M/B \approx B^{-1/2}$ in agreement
with our numerical results.

This same scaling picture can be used to interpret the heat of
fusion at the first-order melting transition at $T_m(B)$.
Assuming that $T_m(B)$ happens to be in the XY critical region, 
we write the free energy
densities below and above $T_m(B)$ as
$-t^2\ln |t|f_s(B|t|^{-4/3})$ and $-t^2\ln |t|f_{\ell}(B|t|^{-4/3})$,
where $f_s$ and $f_{\ell}$ are two different scaling forms for the free
energy density above and below the melting transition.  At the melting
point, these free energy densities must be equal.  Then a little algebra
shows that the jump $\Delta s$
in the entropy density $S = -(\partial f/\partial T)_B$
takes the form
$\Delta s = -t^2\ln |t|(f_{s}^{\prime}(x_m)-f_{\ell}^{\prime}(x_m))$, where
$x_m = B|t_m|^{-4/3}$ is the value of the scaling parameter at the melting
point.  Using $B \sim |t|^{4/3}$ along the melting curve, we find that
along the melting curve
\begin{equation}
\label{dseq}
\Delta s \sim B^{3/2}\ln B.
\end{equation}
Thus the melting transition should have an entropy jump which gets smaller
as the field is reduced.

If $T_\ell(B)$ represents a true phase transition as we desribed using the idea of  
vortex tangles, how can it be understood in terms of the phase coherence?  
One possibility is a {\em line of critical points} for a
continuous phase transition similar to the XY transition in a semi-infinite slab. 
This view provides a natural explanation why the scaling theory with the scaling
variable $B\xi^2$ should be successful. 
Mathematically, one can attach a "phantom" cut-line to the core
of each vortex segment across which phase slips by $2 \pi.$ 
These are benign since continuity and single valuedness of the phase
$\Theta$ at every point is ensured. However, their shape and motion can be
monitored most conveniently to keep track of the  spatial and temporal
disturbance of the phase coherence which have important consequences such as phase
slip dissipation in superconductors.   
For an isolated vortex segment placed at origin
with positive vorticity along
$\hat{z}$, the cut-line may lie straight along the positive $\hat{x}$-axis. A negative
vortex will then have the cut-line on the negative $\hat{x}$-axis. Once we choose a
cut-line for a particular vortex by fixing the reference phase angle, it  provides the
reference for all other vortices. Note that these cut-lines can  only terminate either
at the sample boundary or at the core of vortices of opposite  charges. When there are
several interacting vortices, these cut-lines are no longer straight, and their
tortuosity reflects the phase disturbances induced by deformation of the vortex
configuration away from perfect lattice.  If the vortex lines were
straight, we will observe that the cut-lines associated with each segment 
all line up as we move along a vortex line. Therefore, a cut-line associated with each
vortex line will form a semi-infinite cut-sheet, separated from other
sheets by roughly $L_B\sim B^{-1/2}$. 
As the vortex lines become tortuous in the liquid phase above $T_m$, the cut-sheets
will become wrinkled and our system will look like a three dimensional maze
walled by these sheets.  
Both in the vortex solid and line-liquid phases, this maze will allow a arbitrarily
curved path connecting  both sides(either along $\hat{z}$ or $\hat{x}$-axes) of the
sample, and the average width of the path free of the walls will be
${\cal O}(L_B).$ We conjecture that it is possible that the phase variables may
maintain long range coherence. At low fields(large $L_B$), this tortuous slab contains
many XY phase degrees of freedom which, being confined within the walls, {\em do not
feel} the presence of free vortices, and therefore could conceivably undergo a  phase
transition in the universality class of a zero-field $XY$ model in a ``film" of
thickness 
$\sim L_B$, i.\ e., a quasi 3D-XY transition. This crosses over to a bulk XY transition
as $B \rightarrow 0$. 

There are two possible objections to this picture.  First, our numerical results only
hint at, and certainly do not prove, two separate phase transitions.  Secondly, the
``film'' mentioned above is a dynamical rather than an equilibrium film, in the sense
that its boundaries are not fixed. It is not clear that such a dynamical object could
have an XY phase transition. The boundaries (i.~e.,  cut-sheets of the deformed vortex
lines) are, of course, moving subdiffusively\cite{ryu96,dorsey92} as long as
$\xi_{vz}/d > {\cal O}(10) $. This condition, as we confirmed numerically in section
\ref{vortexsection}, holds true in the range 
$T_m < T < T_\ell$ and makes the above picture more plausible. It also greatly
enhances the chance if we consider pins in real material, since even a single
vortex line then becomes collectively pinned into a glassy state. 

We now discuss a 3D XY-like transition for the infinite slab of
thickness $L_B$.  Such a slab belongs to the $G_2$-class in 
Barber's classification of finite size systems\cite{barber83}).  From this
identification, we can derive many characteristics of the phase transition.
First, consider a thermodynamic quantity for an infinite system in 3D,
which varies as $P_\infty(T) \sim C_\infty t^{-\rho}$, where
$t  = (T-T_c)/T_c$ with $T_c$ the transition temperature for an infinite
system and $\rho$ an appropriate critical exponent.
For a slab of thickness $L_B$, a general finite size scaling
ansatz dictates that
\begin{equation}
\label{ambeq}
P_{L_B}(T) \sim L_B^\omega Q(L_B^\theta \dot{t})
\end{equation}
as $L_B \rightarrow \infty, \dot{t} \rightarrow 0$ with $\theta = 1 / \nu.$
The exponent
$\omega$ is determined by  requiring bulk behavior in the limit $L_B \rightarrow
\infty$; this condition gives $\omega = \rho / \nu.$ 
The transition temperature for a finite ${\bf B}$ is shifted  
\begin{equation}
(T_c - T_c(L_B)) / T_c \sim L_B^{-\lambda}
\end{equation}
and the shift exponent 
$\lambda $ is generally equal to $1 / \nu$, as has been discussed for the
superfluid transition in bulk $^4$He of finite thickness by Ambegaokar 
{\it et al}\cite{amb80}.
For our purposes, it is sufficiently accurate to take 
$\nu  \sim 2/3$, which then agrees very well with  our numerical
result(Eq.~\ref{deltatceq}).  Eq.~\ref{ambeq} needs to be modified for a quantity with
a  logarithmic divergence; it becomes 
$P_\infty (T) \sim C_\infty \ln t$ as $t\rightarrow 0,$ one
modifies
the ansatz\cite{fisher71} so that we have
$P_{L_B}(T) - P_{L_B}(T_0) \sim Q( L_B^\theta \dot{t} ) - Q( L_B^\theta
\dot{t_0} )$, where $T_0$ is some non-critical temperature.  For $\dot{t}
\rightarrow 0$ at a fixed $L_B$, we obtain for such a variable
\begin{equation}
P_{L_B} ( T_c(L_B) ) \sim - C_\infty \theta \ln L_B
\end{equation}
where we have assumed that $Q(z) = {\cal O}(1)$ for $z \rightarrow 0$.
This prediction is in good agreement with the calculated
maximum height of specific heat peak(Eq.~\ref{cpeakeq}),
which for the 3D XY model, has a weak divergence with $\alpha \sim 0.$
Similar results have been discussed for the superfluid transition in He II
of finite thickness by Ambegaokar et al\cite{amb80}.

As the field increases, one may eventually reach the
limit $L_B / \xi_{XY}(T) < 1$, at which the transition at $T_c(L_B)$ will
crossover to a 2D KT universality class and we expect the merging of the two
transitions, $T_\ell = T_m.$ In our model, we believe this happens for a value of $f$
between $1/6$ and $1/18.$ 

\section{Other Recent Simulations}
Towards the completion of this work, we became aware of some of the
more recent studies based on similar models.
We briefly discuss them in comparison with
our main results and interpretations.
To avoid confusion, we use our own conventions for the flux density
given in terms of the frustration $f$ defined earlier and
introduce the anisotropy factor $\Gamma^2 \equiv J_{xy} / J_z$.
For an isotropic system, $\Gamma = 1$ while for ${\rm YBa_2Cu_3O_{7-\delta}}$
it is $\sim {\cal O}(10)$ and for ${\rm Bi_2Sr_2CaCu_2O_8},$ it is
$>> {\cal O}(100).$
We also use the same notation $T_\ell$ for the temperature where $\gamma_{zz}$
drops to zero. Some researchers opted to use $T_z.$

Nguyen and Sudb{\o}\cite{nguyen97} have extended their earlier work
on the anisotropic London loop model\cite{nguyen96}.
Their numerical results in both the vortex structure factor and
$\gamma_{zz}$ for $\Gamma = 1$ with $f=1/32$ follow
a pattern qualitatively similar to our main results for $\Gamma = 1$, $f=1/24$
[See Figure 6 of \cite{nguyen97}] as well as those of Li and
Teitel\cite{li93_1,li93_2}.
By looking at the dependence of $T_\ell(N_z)$ on the thickness of the
system $16 \le N_z \le 96$, and linearly extrapolating the finite size effect,
they conclude that $T_\ell^\infty = T_m$ in the thermodynamic limit($N_z
\rightarrow \infty)$.
Their argument is based on the following observations:1) $T_\ell$ decreases
with an approximately linear dependence on increasing
$N_z$, $T_\ell(N_z+\delta N_z)
\sim T_\ell(N_z) - c \cdot \delta N_z $ with a positive number $c$.
2) In the thermodynamic limit, below the melting transition($T \sim T_m$),
the energy scale for the interlayer phase fluctuation
$T^* \sim \xi^2 J_z$ diverges due to long range lattice order. Therefore,
the linear progression can not continue below  $T_m$.
From these, they conclude that $T_\ell \rightarrow T_m$ in the thermodynamic
limit.

We agree with the validity of the second assumption on general
grounds. However, this does not exclude the possible existence of an intermediate
phase in which the interlayer fluctuation may be suppressed due to the
quasi-long range phase correlations in a line liquid ``phase''.
With this possibility open, $T_m$ is only a lower bound for the $T_\ell.$
It should also be noted that $\gamma_{zz}$ does not show a significant dependence
on size in the region where $0.8 < \gamma_{zz} < 1,$ in temperatures
$T_m < T < 1.8T_m.$ It is only at higher temperatures, near where 
$\gamma_{zz} \rightarrow 0$, that the linear dependence of the shift in $T_\ell$ on
$N_z$ is observable.
In other words, the size dependence of $\gamma_{zz}$ is not trivial as
temperature varies and one should not expect the same size dependence  be uniformly 
applied over the whole temperature range $T_m < T < T_{XY}.$

Furthermore, the linear dependence of shift in $T_\ell$ on $N_z$ in the region where 
$\gamma_{zz} \sim 0$ is
anticipated on more general grounds. In the London limit, $\gamma_{zz}$
measured in the simulation under periodic boundary conditions is
\begin{equation}
\label{lineareq}
\gamma_{zz}(q_x=\pi/\sqrt{N_xN_y}) \sim {J \over V\Gamma^2}
\Big[ 1 - {4J\over N_z \Gamma^2 T \pi} < n_y (q_x=\pi/N_x)
n_y(q+x = -\pi/N_x) > \Big]
\end{equation}
where $n_y(q_x=\pi/N_x)$ is the Fourier component of the vorticity
vector field lying along $\hat{y}$ direction.
Position of $T_\ell$ is governed by the condition that the vortex
fluctuations make the factor in the bracket vanish.
Let us assume that there is a characteristic number of layers $N_z^*$ for which
the true thermodynamic transition is realized at $T_\ell^*.$
For a size $N_z$ smaller than $N_z^*$, $N_z = N_z^* + \delta N_z$($\delta
N_z < 0)$,
linearization of the condition gives
\begin{equation}
T_\ell(N_z) \sim T_\ell^* + g(T_\ell^*) \delta N_z
\end{equation}
with $g(T) = \Gamma^2 \partial [<n_y(\pi/N_x)n_y(-\pi/N_x)>/T ] / \partial T.$
What is the temperature dependence of $<n_y n_y>$? Near $T_m,$ where thermally
activated vortex loops {\em begin} to appear, it is dominated by the vortex
loop fugacity factor and is steeply increasing function of $T$ following an
$S$-shaped curve.
However, near the foot of $\gamma_{zz}$, where our $T_\ell$ is located,
it is numerically observed that
it has reached the plateau and the temperature dependence is dominated
by the $1/T$ factor. It is also consistent with the interpretation of
$T_\ell$ in terms of XY-type unbinding transition.
Since $g(T_\ell) < 0$ and $\delta N_z < 0$ in this region, we then
reach the conclusion that
$T_\ell(N_z)$ linearly increases away from $T_\ell^*$
with $c\dot |N_z^* - N_z|$  as $N_z$ decreases
from the asymptotic(thermodynamic) limit.
Note that we do not assume $T_\ell^* = T_m$ in reaching this conclusion.
If one should follow Nguyen and Sudb{\o} and extrapolate the observation of the  linear
dependence in the limited range of the system size and conclude that
$T_\ell^*(B)=T_m(B)$,
it also follows that we have a paradoxical consequence of predicting
$T_{XY} \rightarrow 0$ for the zero field, based on the similar bahavior numerically
observed for $\gamma_{\alpha\alpha}$ for $f=0.$

Following the convention of Koshelev, we employ the {\em scaled field
variable} $h = \Gamma^2 f$ which characterizes the thermodynamics of the
mixed state as long as $\lambda \rightarrow \infty$.
As we pointed out earlier\cite{ryu97}, $f=1/6$ with
$\Gamma^2 = 1$ ($h = 1/6$) represents a situation where the interlayer
decoupling sets in right at
the melting transition. From our numerical results with $h$ varying from
$1/6$ to
$0$, we believe that there is a universal crossover value of scaled density
$h$ between $1/6$ and $1/18$ which separates a {\em low field} regime from the
{\em high field} regime for which $T_m = T_\ell = T_c(B).$
Koshelev\cite{koshelev97} made a numerical observation that 
$T_\ell \rightarrow
T_m.$ However, it is again made for $\Gamma^2 = J_{xy} / J_z =36, f =
1/36$, i.~e.~
$h = 1$, equivalent to an extremely dense limit.
Recent calculations by Hu and Tachiki\cite{hu97} based on $f=1/25, \Gamma^2
= 10$ in
a larger system size of $50\times 50\times 40$ falls into the same category
with
$h=10/25 \gg 1/6.$ For this dense limit, they observe that $\gamma_{zz}$ drops
sharply to zero at the melting transition, as we had observed earlier for the
$h=1/6$ case in the stacked triangular XY model\cite{ryu97}.

The technical difficulty of simulating the very low field limit $(h < 1/18)$
with large number of field induced vortex lines remains largely unsurmounted.
To minimize the artificial grid pinning effect, one is required to choose a
fairly
small value of $f$( $< 1/32$ for the square grid, $< 1/16$ for the triangular
grid ). However, choice of a large anisotropy factor to mimic HTSC then
tends to
push the model into the high field regime as pointed out earlier.
To access the truly low field regime($B \ll 1 $ tesla for \ybco, $\ll 200 $ gauss  for
\bscco), we had to employ an {\rm isotropic} model for a practical
value of $f < 1/18$. Summarizing this section, we make educated guess on 
the thermodynamics of extremely low field limit, 
where the field-induced vortex degrees
of freedom are no longer viable\cite{tmzero}.  Here, the 
transition at $T_\ell$ takes over the first order melting transition 
as the {\em S-N} transition which possibly belongs
to the universality class of {\em zero field transition of 
XY film} with thickness $a_B$. The phase for $T_m < T < T_\ell$ may still be
considered superconducting in the
sense that a superconducting path, however narrow, may exist across a macroscopic
distance along ${\bf B}$, which defines a tiny, but finite critical current. 
This may be enhanced by collective pinning of the single lines, but resistance
measured with a large driving will show a non-linear IV characteristics and
hysteresis as current has to distribute itself among the superconducting paths and
normal channels.  

The field
induced lines in the low field limit, if we take the world-line analogy, are equivalent
to extremely massive bosons and eventually drop out of the thermodynamics.
They become localized charges which tend to polarize the underlying
vacuum and induce a {\em dielectric breakdown} as the XY medium become
more and
more polarizable as T increases toward $T_\ell^*$.
It is somewhat similar to a  {\em metal-insulator} transition
in a narrow band-gap semiconductor
in which local field gradient(due to the field induced vortices) and
shrinking  band gap
(as $T\rightarrow T_\ell$) conspire to a massive generation of screening
dipole pairs
(vortex loops) leading to a metallic state.

\section{Conclusions}

To summarize, we confirmed the existence of a single first order melting transition at
high vortex densities. 
We also showed that upon melting the local vortex density increases due to the
screening effect of thermally generated vortex loops. 
More significantly, however, we
observed that,  in the absence of
disorder, destruction of phase coherence in a superconductor may proceed by two separate
transitions at low magnetic fields: a quasi-long range phase coherence parallel to the
field disappears at a temperature $T_\ell$ higher than $T_m$ at which the lattice
periodicity disappears and true long range phase coherence is lost. In this low-field
regime,
the lattice first melts into a liquid of lines  with a finite entanglement
length
along the applied field.   These lines eventually disappear through increasing
entanglement, and through their interaction with thermally induced vortex and
antivortex  loops. 
While the melting transition is best characterized by the disappearance
of Bragg peaks for the vortex lines and a delta function peak in the specific heat,
there is a narrow region above
$T_m$  where we observe dramatic changes in
dissipation tensor which coincide with jump in the local vortex density and
disappearance of the longitudinal phase  rigidity, $\gamma_{zz}=0$.
Instead of being a gradual crossover, we propose that a possible transition at 
$T_\ell \ne T_m$ sets in at low densities.  
It tracks the broad peak in specific heat as B increases, obeying the
behavior of 3D zero field XY system confined to a semi-infinite slab of
finite thickness $L_B \sim B^{-1/2}.$
It can alternatively described in terms of appearence of connected vortex tangle which 
effectively leads to decoupling of neighbouring layers. 
Within this picture, origins of several puzzling and conflicting anomalies recently
obtained on \bscco and \ybco may be understood.

\section{Acknowledgments}
This work has been supported by the Midwest
Superconductivity Consortium at Purdue University, through Grant
DE-FG02-90ER45427.    
Calculations were carried out, in part, on the network
of SP-2 computers of the Ohio Supercomputer Center.

\break

\break

\break

\break

\break

\break

\break

\break

\break

\break

\break

\break

\break

\break

\break

\break

\break

\break

\break

\break

\break

\break

\break

\break

\break

\break


\begin{thebibliography}{10}
\bibitem{sharpjump}For example, see Fig.\  1 of \cite{kwok92}.
The jump is most pronounced for $1.5 < B < 4$ tesla.
\bibitem{errata} In both \cite{sasik95} and \cite{ryu96},
the contribution to the
total internal energy coming from the temperature dependence of the
coupling strengths of the effective free energy functional has been neglected.
leading to an underestimate of the value for the heat of melting.
When this temperature dependence is taken into account,
the results of \cite{ryu96}
give $\triangle S \sim 0.28 k_B$. Updated
LLL results also have a similar value.
\bibitem{tmzero} For a hypothetical isotropic superconductor with $d \sim
\xi \sim 100
\AA$,
${\rm L}_x \times {\rm L}_y \times {\rm L}_z \sim 0.1\times 0.1 \times 10^{-3}{\rm cm}^3,$ we
have about
${\cal N}_\phi \sim 10^{13}$ ``grains" of the XY model.  For a given field
$B$(in gauss),
the number of  pancake vortex
degrees of freedom is
${\cal N}^{2D}_v \sim 10^{-5} B / (10^{-6} \phi_0)
\sim {\cal O}(10^{8}) \times B$. This number will begin to match that of XY
degrees of
freedom if $B \sim 10 {\rm tesla}.$ Note that if the lattice melts into a
line liquid with
an infinite correlation length along $\hat{z}$, the relevant degrees of
freedom right at
$T_m$ will be ${\cal N}^{3D}_v \sim 10^{-2} B /  \phi_0 \sim 10^5 B.$
For the speculative picture of $T_\ell$ as the XY transition of a thick film, a
crossover from the ``finite slab" to the 2D XY regime should occur when
$a_B \sim {\cal O}(\xi).$ This is met when $ B \sim 10 {\rm tesla}.$
For this hypothetical sample, one may predict that for $B > 10 {\rm
tesla}$, the
SN transition is dominated by the vortex degrees of freedom
and a separate transition at $T_\ell$ is pre-empted by the first order
melting transition,
since  the lattice melts into a 2D-slab regime directly.
On the other hand, as $B$ decreases below $10$ tesla toward zero, the
melting transition
should follow a finite size scaling behavior as the vortex ``line" degrees
of freedom falls
below that of thermodynamic limit, while the XY degrees of freedom takes
charge and
drive the transition at $T_\ell$ of the thick, but tortuous XY slab of
finite thickness $\sim
a_B$.
\end{thebibliography}
\end{document}